\documentclass[a4paper,11pt]{article}
\pdfoutput=1 

\usepackage{jheppub} 

\usepackage[T1]{fontenc} 
\usepackage{units}
\usepackage{subcaption}
\usepackage{graphicx}
\usepackage{float}
\RequirePackage[dvipsnames]{xcolor}
\usepackage{slashed}

\newcommand{\Exp}[1]{\exp{\left\{#1\right\}}}

\title{\boldmath  Collinear spin correlations of final-state radiation in dense QCD matter}

\newcommand{\nn}{\nonumber\\ }

\def\k{{\boldsymbol k}}
\def\q{{\boldsymbol q}}
\def\p{{\boldsymbol p}}

\def\l{{\boldsymbol l}}

\def\x{{\boldsymbol x}}
\def\y{{\boldsymbol y}}

\def\r{{\boldsymbol r}}
\def\v{{\boldsymbol v}}
\def\u{{\boldsymbol u}}

\def\bkappa{{\boldsymbol \kappa}}
\def\bepsilon{{\boldsymbol \epsilon}}

\def\Re{\text{Re} \, }
\def\Im{\text{Im}}
\def\Tr{\text{Tr} \,}

\newcommand{\cK}{\mathcal{K}}
\newcommand{\cM}{\mathcal{M}}
\newcommand{\cP}{\mathcal{P}}

\newcommand{\cA}{\mathcal{A}}
\newcommand{\cB}{\mathcal{B}}
\newcommand{\cC}{\mathcal{C}}
\newcommand{\cD}{\mathcal{D}}

\newcommand{\cQ}{\mathcal{Q}}

\author[a,b,c]{Jo\~ao M. Silva,}
\author[a]{Alba Soto-Ontoso}
\affiliation[a]{Departamento de Física Teórica y del Cosmos, Universidad de Granada, Campus de Fuentenueva, E-18071 Granada, Spain}
\affiliation[b]{Laboratório de Instrumentação e Física Experimental de Partículas (LIP), Av. Prof. Gama Pinto, 2, 1649-003 Lisbon, Portugal}
\affiliation[c]{Departamento de Física, Instituto Superior Técnico (IST), Universidade de Lisboa, Av. Rovisco Pais 1, 1049-001 Lisbon, Portugal}

\emailAdd{joao.m.da.silva@tecnico.ulisboa.pt}
\emailAdd{aontoso@ugr.es}

\abstract{
Spin correlations are required to reproduce the correct azimuthal dependence of matrix elements for successive branchings at disparate angles in QCD jets. In this paper, we study modifications to this, $\cos(2\psi_{12})$, azimuthal pattern in the presence of a quark-gluon plasma. To that end, we consider a simplified setup in which a narrow and energetic QCD antenna is formed inside a medium of fixed length and radiates a collinear emission outside it. The calculation includes both light and heavy-quarks. Further, we do not include medium-induced spin-flip interactions since they are energy suppressed in our formalism.
We show that the amplitude of the azimuthal modulation in the presence of a medium is always suppressed with respect to the vacuum baseline, with its magnitude depending on the medium properties and splitting kinematics. 
For a medium with a momentum space anisotropy, we find that the azimuthal modulation acquires a phase shift, i.e., $\cos(2\psi_{12}) \to \cos(2\psi_{12}+\phi_{\rm med
})$, where $\phi_{\rm med
}$ is a process-dependent function that again depends on the medium properties and splitting kinematics. This work provides theory guidance for implementing spin-driven interference effects in phenomenological studies of jet quenching in heavy-ion collisions.
}

\begin{document} 
\maketitle
\flushbottom

\section{Introduction}
\label{sec:intro}
The calculation of theoretical predictions for jet observables at collider experiments involves $n$-parton matrix-elements. QCD multiparton matrix elements adopt a particularly simple form in a variety of ordered limits, where kinematic properties of the emissions, such as their energy or angle, decrease from one emission to the next~\cite{Bassetto:1983mvz,Dokshitzer:1992ip,Campbell:1997hg,Catani:1999ss}. For instance, when all emissions are both soft and collinear, and widely separated in rapidity from each other, the tree-level $n$-parton matrix-element reduces to the product of $n$ single-emission matrix-elements. This factorisation property lies at the core of parton shower algorithms~\cite{Campbell:2022qmc}. 

Quantum interference effects driven by colour and spin degrees of freedom break this independent-emission picture. Colour interferences are
relevant when considering multiple emissions at commensurate angles and strongly-ordered in energies. They lead to a suppression of soft, wide-angle emissions, i.e., the phenomenon of angular ordering of soft radiation~\cite{Dokshitzer:1982fh,Webber:1983if,Marchesini:1987cf,Dokshitzer:1987nm,Gustafson:1987rq,Nagy:2019pjp,Hamilton:2020rcu,Forshaw:2025fif}.
Spin correlations appear when considering processes with intermediate off-shell gluons and become relevant for soft, wide-angle gluons that split collinearly or in successive collinear emissions strongly ordered in angle. In this work we focus on the latter.
%
These interference effects produce a non-trivial azimuthal correlation between the planes of consecutive emissions~\cite{Efremov:1981sh,Shatz:1983hv,Webber:1986mc,Webber:1987uy,Collins:1987cp,Knowles:1987cu,Knowles:1988vs,Knowles:1988hu,Richardson:2018pvo,Forshaw:2019ver,Corcella:2000bw,Karlberg:2021kwr,Hamilton:2021dyz,Bewick:2023tfi,Hoche:2025anb}. 

 
This paper studies spin-driven interference effects between successive QCD splittings when jet fragmentation evolves in a quark-gluon plasma (QGP)~\cite{Majumder:2010qh,Mehtar-Tani:2013pia,Blaizot:2015lma,Connors:2017ptx,Cunqueiro:2021wls,Apolinario:2022vzg,Apolinario:2024equ,Wang:2025lct}. We will assume a weak-coupling description of the QGP, justified at high enough temperatures, so that perturbative QCD techniques, in particular the BDMPS-Z formalism~\cite{Baier:1996kr,Zakharov:1996fv,Wiedemann:2000za}, can be applied. The medium is then modeled as a stochastic, background colour field. This classical field contains information about both the colour density of the medium and the scattering potential that controls the interaction between the traversing partons and the medium. Considering all partons to be much more energetic than the relevant medium scales, spin-flip interactions are suppressed and thus the main effect of the medium is to colour-rotate the parton and to modify its transverse momentum distribution. The latter can be directional-dependent for some phenomenologically relevant scenarios in heavy-ion collisions. Examples of such cases are: strong longitudinal polarized fields in the glasma phase~\cite{Ipp:2020mjc,Carrington:2021dvw,Barata:2024xwy, Avramescu:2024poa,Avramescu:2024xts}, anisotropic phase-space distributions in the pre-equilibrium phase~\cite{Boguslavski:2024ezg,Boguslavski:2024jwr,Altenburger:2025iqa,Barata:2025agq} or matter gradients and flow in non-equilibrium QGP~\cite{Sadofyev:2021ohn,Antiporda:2021hpk,Barata:2022krd,Andres:2022ndd,Fu:2022idl,Barata:2023qds,Barata:2023zqg, Kuzmin:2023hko,Kuzmin:2024smy,Ke:2024emw}. In this work, we will explore the case of a medium with a momentum space anisotropy for which rotational invariance is broken, while keeping the assumption of homogeneity~\cite{Hauksson:2021okc,Hauksson:2023tze,Barata:2024bqp, Barata:2025uxp}.



The case of colour interferences has been studied in a series of works at the level of two-emissions using the antenna setup, where a first, energetic splitting then emits a soft gluon, either inside or outside the medium~\cite{Mehtar-Tani:2010ebp,Mehtar-Tani:2011hma,Casalderrey-Solana:2011ule,Mehtar-Tani:2011vlz,Mehtar-Tani:2011lic,Armesto:2011ir,Mehtar-Tani:2012mfa,Casalderrey-Solana:2012evi,Calvo:2014cba,Barata:2021byj,Abreu:2024wka,Andres:2025prc,Kuzmin:2025fyu}. It has been observed that medium-interactions during both the antenna formation and its propagation modify the rate and phase-space of soft gluon emissions with respect to vacuum. In particular, colour exchanges between the antenna legs and the medium degrade the colour-correlation of the antenna and, eventually, lead to a fully decoherent state in which each of the legs radiate independently. Consequently, the regime where the two-emissions can be considered independent and thus the two-body matrix-element factorizes is actually enhanced with respect to vacuum. 

On the other hand, the case of spin-driven interferences has never been explored. In this paper we compute the fully differential cross section for two collinear and strongly angular-ordered emissions. For simplicity, we take the second splitting to be outside of the medium, as the calculation of the full $1\rightarrow 3$ matrix-element inside a QCD medium is considerably more involved~\cite{Fickinger:2013xwa,Casalderrey-Solana:2015bww,Arnold:2015qya,Arnold:2022mby,Arnold:2023qwi,Arnold:2024whj}. Spin correlations between successive splittings only arise when there is an intermediate off-shell gluon,
as a consequence of summing coherently over the spin states of this intermediate particle. Therefore, we restrict the calculation to processes like $a \rightarrow b\, g \rightarrow b\,c\,d$, where $b,c,d$ can be either light quarks ($q$), heavy-quarks ($Q$) or gluons ($g$), i.e., $a,b,c,d\in (q, Q, g)$. In vacuum, the fully differential cross-section for such a configuration can be schematically written as 
\begin{align}
\label{eq1:vac_xs}
	(2\pi)^2\frac{d\sigma^{\rm vac}}{\,dz_1dz_2d\theta_{1}d\theta_{2}d\phi_{1}d\phi_2} = \frac{d\sigma^{\rm vac}}{dz_1d\theta_{1}}\frac{d\sigma^{\rm vac}}{dz_2 d\theta_{2}}\left[1 + a\cos(2\psi_{12})\right]\,,
\end{align}
where $z_{1(2)}$ and $\theta_{1(2)}$ are the energy sharing fraction and the opening angle of the first (second) splitting, respectively. The deviation from the iterated emission picture is captured by the term in brackets in Eq.~\eqref{eq1:vac_xs}, where the coefficient $a(z_1,z_2,\theta_1,\theta_2)$ depends on the final-state under consideration and its kinematics and $\psi_{12} = \phi_1-\phi_2$ is the azimuthal difference between
the planes spanned by the two branchings. We will come back to the exact definition of azimuthal angles below. The phenomenological implications of the azimuthal modulation in Eq.~\eqref{eq1:vac_xs} have been discussed in, for instance, Refs.~\cite{Richardson:2018pvo,Chen:2020adz,Karlberg:2021kwr,Hamilton:2021dyz,Hoche:2025anb,Song:2025bdj}.  

The main result of this paper is that for a dense, anisotropic QCD medium the fully-differential cross-section reads
\begin{align}
\label{eq2:med_xs}
	(2\pi)^2\frac{d\sigma^{\rm med}}{\,dz_1dz_2d\theta_{1}d\theta_{2}d\phi_{1}d\phi_2} = \frac{d\sigma^{\rm vac}}{dz_1d\theta_{1}}\frac{d\sigma^{\rm vac}}{dz_2 d\theta_{2}}(1+F_{\rm med})\left[1 + a_{\rm med}\cos(2\psi_{12}+\phi_{\rm med})\right]\,,
\end{align}
where, in the vacuum limit, $1+F_{\rm med}\to 1$, $a_{{\rm med}}\to a$ and $\phi_{\rm med}\to 0$ so that Eq.~\eqref{eq2:med_xs} reduces to Eq.~\eqref{eq1:vac_xs}. Medium modifications enter in three terms in the previous equation. First, $F_{\rm med}(E_a,z_1,\theta_1, \phi_1)$ is the same medium modification factor as the one introduced in Refs.~\cite{Dominguez:2019ges,Isaksen:2020npj} and controls the nuclear modification to the antenna-formation cross-section. Second, $a_{\rm med}=a(z_1,z_2,\theta_1,\theta_2)\, f(E_a, z_1, \theta_1, \phi_1)$ suppresses the azimuthal modulation with respect to vacuum by a factor ($f < 1$) that depends on the phase-space point and the medium properties. Finally, medium interactions induce a phase-shift encoded by $\phi_{\rm med}$. In the isotropic case, since there is no preferred direction, this phase-shift vanishes together with the $\phi_1$ dependence in $a_{\rm med}$. We highlight that Eq.~\eqref{eq2:med_xs} only holds when the second splitting happens outside the medium. If that were not the case both the factorised structure would be broken and the $\psi_{12}$-dependence would be more complex. 


The rest of this paper is organized as follows. In Section~\ref{sec:vac_spin_corr}, we discuss the regime of validity of our calculation and revisit the vacuum result. The differential cross-section for two-successive splittings including medium effects during the antenna formation and propagation is presented in Sec.~\ref{sec:in_medium_spin_corr}. In Sec.~\ref{sec:numerics}, we present a numerical evaluation of the formulas obtained in \ref{sec:in_medium_spin_corr} and we identify the regions of phase-space where medium effects are most relevant. We conclude in Sec.~\ref{sec:conclusions} with a summary and a discussion of the steps required to transform the present calculation into a phenomenological tool. Technical details on the calculation are reported in Appendix~\ref{app:details}.

\section{Overview of collinear spin correlations in vacuum}\label{sec:vac_spin_corr}

\subsection{Heuristic discussion}
Before delving into the actual calculation, let us explain the kinematic regime in which the calculation is done and qualitatively discuss the emergence of spin-driven azimuthal correlations between subsequent splittings.

We consider $1 \to 3$ processes such as $a \to b g \to b c d$, i.e., any process
with final-state partons $b$, $c$, $d$ which includes an intermediate off-shell gluon that splits as
$g \to c d$. 
We are interested in the limit where all final-state particles are \textit{quasi}-collinear~\cite{Keller:1998tf,Catani:2000ef}.\footnote{This limit is a generalisation of the usual collinear limit~\cite{Catani:1999ss}, allowing for the definition of universal splitting functions that take mass effects into account~\cite{Campbell:1997hg,Catani:2000ef,Dhani:2023uxu,Craft:2023aew}. Put simply, one defines such a limit by taking all relative transverse momenta (or, equivalently, pair-wise angles) and on-shell masses to be small, while keeping their ratio fixed.} Further, we shall consider this limit in the strongly-ordered regime where one enforces
\begin{equation}\label{eq:ang_ordering}
\theta_{2} \ll  \theta_{1} \ll 1\,, \quad (\theta_{1}\equiv \theta_{bg}, \theta_{2}\equiv \theta_{cd}) \,
\end{equation}
being $\theta_{ij}$ the opening angle of the $ij$ pair of particles.
In this strongly-ordered limit, the scattering amplitude for the $a\to b\,g \to b\,c\,d$ process factorizes into the production of a nearly on-shell gluon ($\cM_1$) and its subsequent branching ($\cM_2$).\footnote{Strictly speaking, one also needs the off-shell gluon to not take all the energy in $a\rightarrow b \,g$ for massless $a$ and $b$ and the splitting $g\rightarrow c\, d$ to not be highly asymmetric in energy for massive $c$ and $d$. See e.g. Refs.~\cite{Campbell:1997hg, Somogyi:2005xz, Braun-White:2022rtg,Craft:2023aew} for a more careful treatment of strongly ordered/"iterated" collinear limits.} That is, the colour-stripped amplitude reads
\begin{equation}\label{eq:general_factorised_M}
	i\cM_{a\to cbd} \propto \sum_{\lambda_g,\lambda_a} \cM_{2}^{\lambda_g \lambda_c \lambda_d}(p_c, p_d)\cM_{1}^{\lambda_{a} \lambda_{g}\lambda_b}(p_b, p_g)\cM_0^{\lambda_a}(p_a)\, ,
\end{equation}
where $\cM_0$ is some initial production amplitude, $p_a = p_b+p_c+p_d$, $p_g = p_c+p_d$ and each $\cM_i$ is fully contracted in Dirac space and involves the propagator pole of the corresponding $i$-th splitting. The colour indices have been omitted for clarity and $\lambda_{a,b,c,d}$ are polarisation indices.
The crucial point in Eq.~\eqref{eq:general_factorised_M} is that, despite $\cM_1$ and $\cM_2$ being factorised in Lorentz and Dirac indices, there is a coherent sum over physical polarisations $\lambda_g$, which upon squaring gives a double sum over polarisations, $\lambda_g$ and $\lambda_g'$. The interference terms, i.e., when $\lambda_g \neq \lambda_g'$, result in a dependence on the relative azimuthal angle between the planes of the two consecutive splittings $\psi_{12}\equiv \phi_2-\phi_1$ (see Eq.~\eqref{eq:phi_to_psi}). 

To see this more clearly, we note that the dependence on azimuthal angles can be trivially factored out from each amplitude $\cM_i$ (see Appendix~\ref{app:vertex}), such that the amplitude can be written as
%
\begin{align}\label{eq:M_decomp}
	i\cM_{a\to bg \to bcd} &\propto \sum_{\lambda_g}\tilde \cM_2^{\lambda_g\lambda_c\lambda_d}(z_2, |\q_{2}|)\tilde \cM_1^{\lambda_a\lambda_g \lambda_b}(z_1, |\q_{1}|)\cM_0(p_a)  \nonumber \\
    &\times e^{i(\lambda_g-\bar \lambda_c- \bar \lambda_d) \phi_2} e^{i(\bar\lambda_a-\lambda_g - \bar\lambda_b) \phi_1}\,,
\end{align}
where the barred polarisation indices are given by $\bar\lambda=\lambda$ if the particle is a gluon, and $\bar\lambda=\pm \lambda/2$ if the particle is a quark or an anti-quark, respectively. The light-cone energy fractions $z_i$ and the relative transverse momenta of each splitting $\q_i$ are defined in Eq.~\eqref{eq:relative_momenta_def} and the azimuthal angle of each splitting is defined as $\phi_i=\tan^{-1}(\q^y_i/\q^x_i)$. From Eq.~\eqref{eq:M_decomp}, it is clear that, upon squaring the amplitude to obtain the cross-section, all azimuthal angle phases vanish except those proportional to $\lambda_g$, the intermediate gluon's polarisation state. Thus, the amplitude squared reads\footnote{We assume that the initial parton is unpolarized, i.e., that $\cM_0$ is spin independent. In doing so, we are discarding potential azimuthal correlations with initial-state particles.}
%
\begin{align}\label{eq:M2_decomp}
	|\cM_{a\to bg \to bcd}|^2 \propto |\cM_0(p_a)|^2\sum_{\lambda_g, \lambda_g'} F^{\lambda_g \lambda_g'}(z_i,|\q_i|)e^{i(\lambda_g-\lambda_g')\psi_{12}}\,,
\end{align}
where  
\begin{align}\label{eq:F_definition}
	F^{\lambda_g \lambda_g'}(z_i, |\q_i|) & = \frac{1}{2}\sum_{\lambda_a, \lambda_b,\lambda_c,\lambda_d}\tilde \cM_2^{\lambda_g \lambda_c \lambda_d}(z_2,|\q_{2}|)\tilde \cM_1^{\lambda_a \lambda_g \lambda_b}(z_1,|\q_{1}|)\nonumber\\
    &\times \left(\tilde \cM_2^{\lambda_g' \lambda_c \lambda_d}(z_2,|\q_{2}|)\tilde \cM_1^{\lambda_a \lambda_g' \lambda_b}(z_1,|\q_{1}|)\right)^{\ast}\,.
\end{align} 
%
Since $|\cM|^2$ must be real, by defining $F^{\lambda_g\lambda_g'} = F_1^{\lambda_g} \delta^{{\lambda_g},{\lambda'_g}} + F_2^{\lambda_g}\delta^{\lambda_g, -\lambda'_g}$ one can write
\begin{align}\label{eq:real_imag_F_M2}
	|\cM_{a\to bcd}|^2 \propto  |\cM_0(p_a)|^2 &\times \sum_{\lambda_g} \left[\Re F_1^{\lambda_g}(|\q_{i}|)+ \Re F_2^{\lambda_g}(|\q_{i}|)\cos(2\psi_{12}) \right. \nonumber\\
    & -\left. \lambda_g\,\Im  \,F_2^{\lambda_g}(|\q_{i}|)\sin(2\psi_{12})\right]\,.
\end{align}
At this stage we have isolated the azimuthal dependence into two terms proportional to $\cos (2\psi_{12})$ and $\sin (2\psi_{12})$. This dependence is simply a consequence of coherently summing over gluon polarisation states, since it comes from the terms with $\lambda_g' \neq \lambda_g$. In vacuum, the functions $F_i^{\lambda_g}$ do not have an explicit polarisation dependence, i.e.,  $F_i^{\lambda_g} = F_i$. Thus, the $\sin(2\psi_{12})$ term vanishes and we are left with a $\cos(2\psi_{12})$ modulation which is well known in the literature~\cite{Shatz:1983hv, Knowles:1987cu,Collins:1987cp}. The phenomenological implications of this modulation have recently been studied in e.g.~\cite{Richardson:2018pvo,Chen:2020adz,Karlberg:2021kwr,Hamilton:2021dyz,Hoche:2025anb,Song:2025bdj}.
%

\subsection{Cross-section computation}\label{subsec:vac_xsec}
Let us now explicitly calculate the azimuthal angle modulation of the vacuum cross-section for the process $a\rightarrow b \, g\rightarrow b \, c \, d$ in the strongly-ordered \textit{quasi}-collinear limit. As already mentioned, in this limit the amplitude can be written as a product of propagator denominators and vertices that are fully contracted in Dirac and Lorentz indices
\begin{equation}\label{eq:vac_M}
    i\cM_{a\to bcd} = \sum_{\lambda_g}\sum_{m,l} \left(\frac{i}{Q_{g}^2}igT_{ij}^m V_{g\rightarrow c d}^{\lambda_g \lambda_c \lambda_d}(z_2,\q_{2})\right)\left(\frac{i}{Q_a^2-m_1^2}  igT^m_{kl}V_{a\rightarrow g b}^{\lambda_a\lambda_g \lambda_b}(z_1,\q_{1})\right)\cM_0^{l}(p_a)\,,
\end{equation}
where we explicitly sum over the colour and polarisation of the intermediate gluon and the colour of the initial parton $a$. For $q\rightarrow gq$ the colour matrices are in the fundamental representation $T^m_{kl} = t^m_{kl}$ while for $g\rightarrow gg$ one has the adjoint representation $T^m_{kl} = -if^{mkl}$. The vertices $V$ are defined and calculated in Appendix~\ref{app:vertex} and the propagator denominators are written in the collinear limit as a function of the virtuality  
\begin{subequations}\label{eq:ordering:vacuum}
\begin{align}
	& Q_a^2-m_1^2 = p_a^2-m_a^2= \frac{\q_1^2+z_1^2 m_1^2}{z_1(1-z_1)}+\frac{\q_2^2 + m_2^2}{z_1z_2(1-z_2)} \approx \frac{\q_1^2+z_1^2 m_1^2}{z_1(1-z_1)}\label{eq:qa-m12}\,,\\
	& Q_g^2 = (p_c + p_d)^2 = \frac{\q_2^2 + m_2^2}{z_2(1-z_2)}\,,
\end{align}
\end{subequations}
where $m_i$ is the quark mass in the $i$-th splitting and in Eq.~\eqref{eq:qa-m12} we imposed the strong ordering condition $\theta_1\gg \theta_2$.  
Using the explicit vertex expressions in Appendix~\ref{app:vertex}, the squared matrix element summed over quantum numbers can be straightforwardly computed. To calculate the differential cross-section we then multiply by the appropriate phase-space element (see Eq.~\eqref{eq:phase-space-final}), resulting in
\begin{equation}\label{eq:xsection-vac}
	 (2\pi)^2\frac{d\sigma^{\rm vac}}{\sigma_0\,d^6\Omega_{\rm PS}} = \left(\frac{\alpha_s}{\pi}\right)^2 C_1 C_2\frac{P^M_{a\rightarrow bg}(z_1,\theta_{1})}{\theta_1(1+\tilde\theta^2_{m,1})}\frac{ P^M_{g\rightarrow cd}(z_2,\theta_{2})}{\theta_2(1+\tilde\theta^2_{m,2})}[1 + a \cos(2\psi_{12})]\,,
\end{equation}
which has the structure of Eq.~\eqref{eq1:vac_xs} with
\begin{equation}
\label{eq:adef-vac}
a \equiv S_{g\rightarrow cd}\frac{4z_2(1-z_2)(1-z_1)}{z_1[P_{a\rightarrow b g}(z_1)+z_1\tilde\theta^2_{m,1}][P_{g\rightarrow cd}(z_2)+\tilde\theta^2_{m,2}]}\, .
\end{equation}
%
The sign of $a$ indicates whether spin correlations are enhanced when the two splittings occur in the same plane ($a>0$) or in perpendicular planes ($a<0$). The factor $C_i$ is the colour factor associated to the $i$-th splitting, $S_{g\rightarrow gg} = +1$ and $S_{g\rightarrow q\bar q} = -1$. The functions $P^M_{i\rightarrow jk}$ are the spin-averaged massive splitting functions that read 
\begin{subequations}\label{eq:mass-ap}
\begin{align}
    & P^{M}_{a\rightarrow bg}(z_1,\theta_{1}) = \frac{\theta_{1}^2 P_{a\rightarrow gb}(z_1)+z_1\theta_{m,1}^2}{\theta_{1}^2 + \theta_{m,1}^2 }\, ,\\
    & P^M_{g\rightarrow cd}(z_2,\theta_{2}) = \frac{\theta_{2}^2P_{g\rightarrow cd}(z_2) + \theta_{m,2}^2}{\theta_{2}^2+\theta_{m,2}^2}\,,
\end{align}
\end{subequations}
with $P_{i\rightarrow jk}$ given by Eq.~\eqref{eq:vac_splitting_functions} and we have introduced dead-cone angles
\begin{align}
\label{eq:deadcone-def}
	& \theta_{m,1}  = \frac{m_1}{(1-z_1)E_a}\,,\qquad \theta_{m,2} = \frac{m_2}{z_1z_2(1-z_2)E_a}\, ,
\end{align}
and their ratios with respect to the opening angle of the splitting, i.e., $\tilde\theta_{m,i} \equiv \theta_{m,i}/\theta_i$. In Eq.~\eqref{eq:deadcone-def}, $E_a = p_a^+/\sqrt{2}$ is the energy of particle $a$. 
Finally, the initial production cross-section (see Eq.~\eqref{eq:phase-space-final}) reads
\begin{equation}\label{eq:sigma_0_def}
	\sigma_0 =  \sum_{l}\int d^3\Omega_a|\cM_0^{l}(p_a)|^2.
\end{equation}
Note that after integrating Eq.~\eqref{eq:xsection-vac} over $\phi_1$ and $\psi_{12}$ we simply recover the product of the $a\rightarrow b\,g$ and $g\rightarrow c\,d$ massive splitting functions.

\begin{figure}
    \centering
         \includegraphics[width=0.49\textwidth]{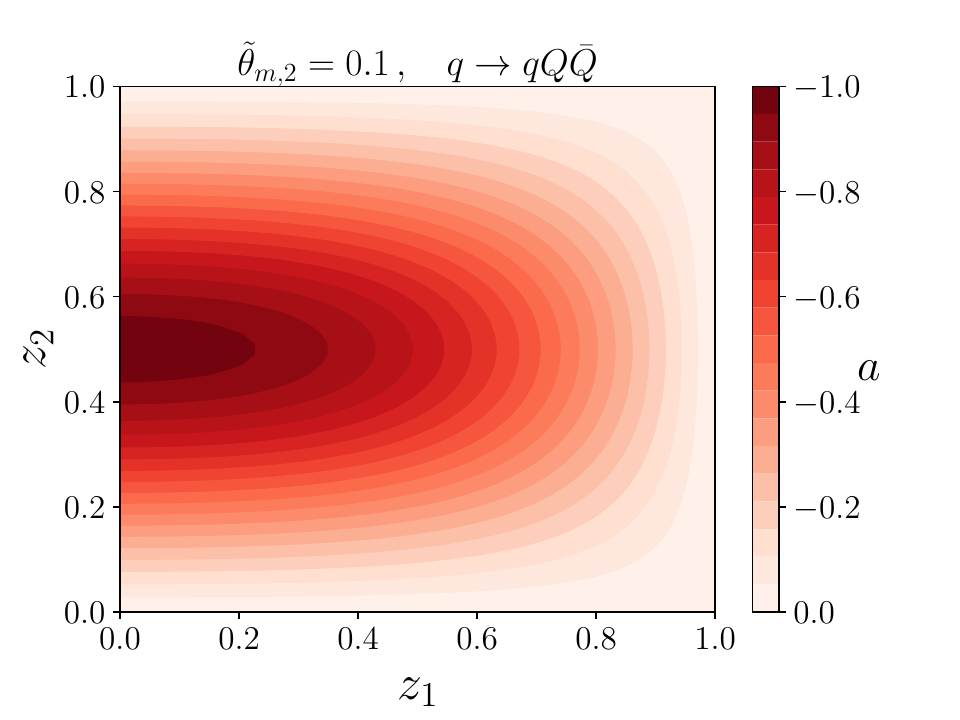}
         \includegraphics[width=0.49\textwidth]{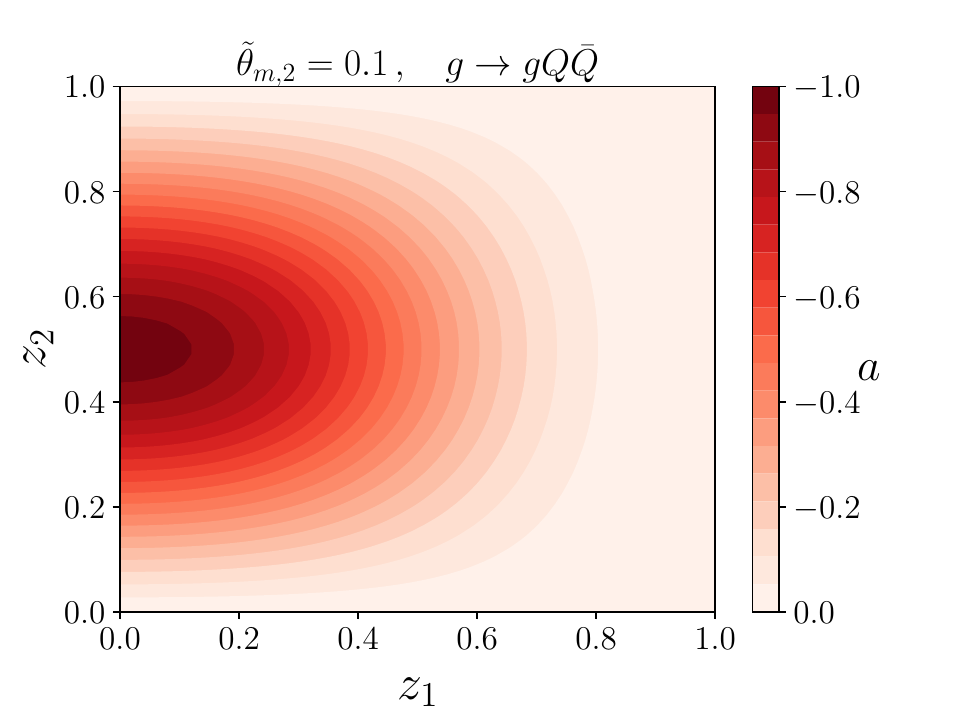} 
         \\
        \includegraphics[width=0.49\textwidth]{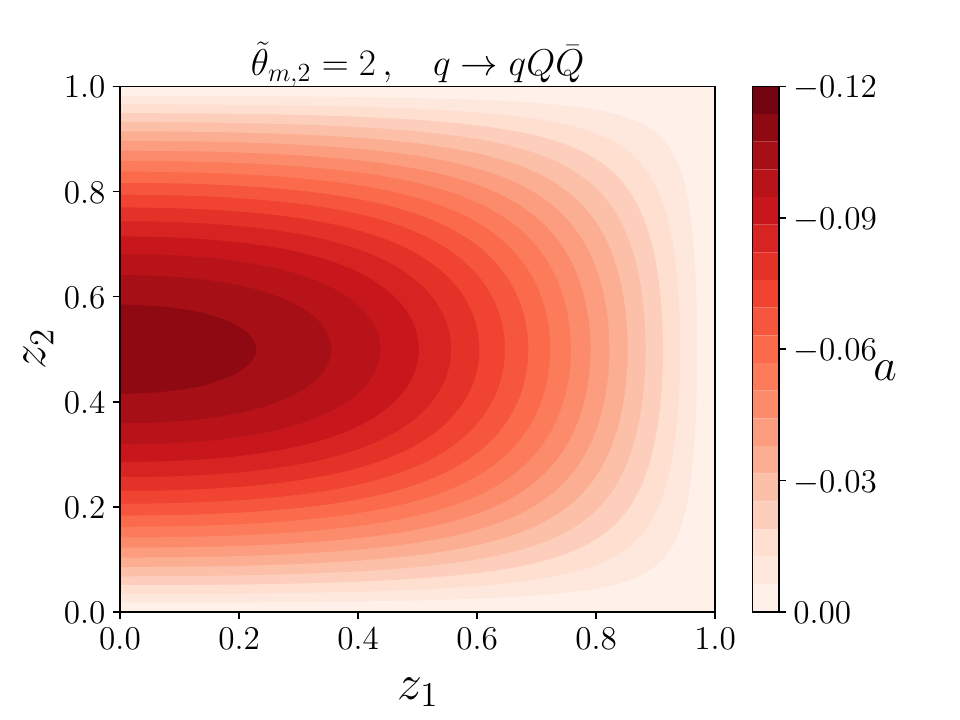}
         \includegraphics[width=0.49\textwidth]{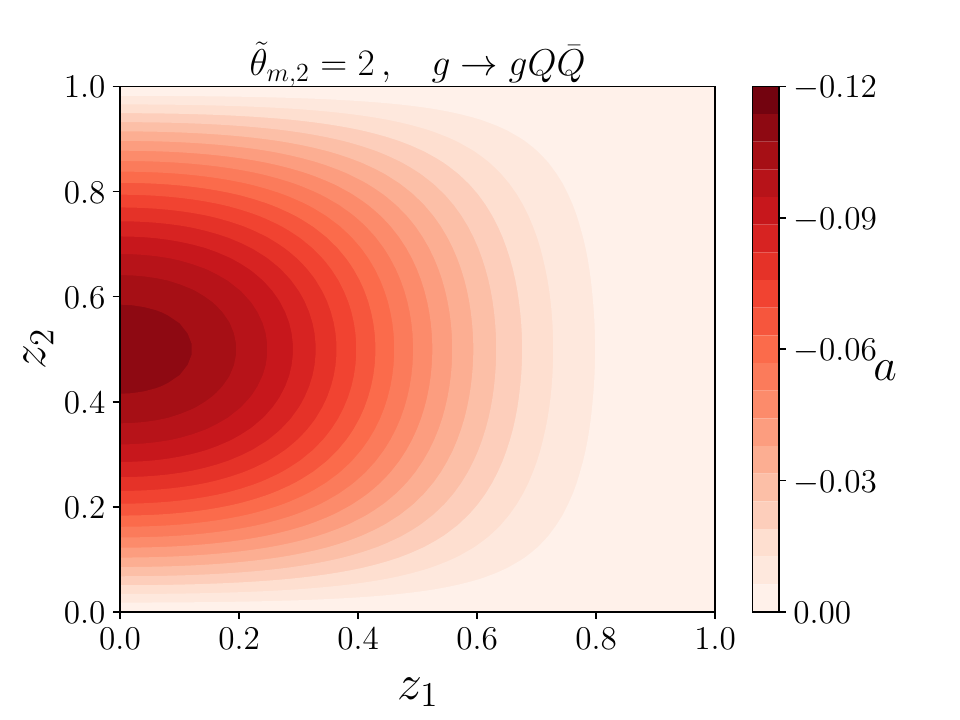} 
     \caption{Amplitude of the spin modulation signal, $a$, defined in Eq.~\eqref{eq:adef-vac}, as a function of both energy fractions $z_i$ for two different first splittings (massless $q\rightarrow qg$ on the left and $g\rightarrow gg$ on the right), fixing the second splitting to be $g\rightarrow Q\bar Q$. From top to bottom we change the opening angle of the second splitting with respect to the dead-cone angle, i.e., $\tilde\theta_{m,2}=\theta_{m,2}/\theta_2$ with $\theta_{m,2}$ defined in Eq.~\eqref{eq:deadcone-def}.} 
     \label{fig:a_vac}
\end{figure}

We evaluate the amplitude of the spin-modulation signal, i.e., Eq.~\eqref{eq:adef-vac}, as a function of ($z_1,z_2$) in Fig.~\ref{fig:a_vac}. In these results, $\tilde\theta_{m,1}=0$. The left and right panels of Fig.~\ref{fig:a_vac} contain the $q\to qg \to q q'\bar q'$ and $g\to gg \to g q\bar q$ processes, respectively. The other relevant scale in Eq~\eqref{eq:adef-vac} is $\tilde\theta_{m,2}$. Since the massless case has been recently discussed~\cite{Richardson:2018pvo,Karlberg:2021kwr,Hoche:2025anb}, we focus on a massive $g\to Q\bar Q$ second splitting and choose $\tilde\theta_{m,2}=0.1$ (top panels) and $\tilde\theta_{m,2}=2$ (bottom panels). 
Let us first examine the top row where $\tilde\theta_{m,2}=0.1$ and thus mass effects are suppressed. For both channels, we observe that the size of spin-correlations reaches its maximum around $z_1\lesssim 0.1$ and $z_2\sim0.5$, i.e., when the initial parton is soft and the energy is shared democratically in the second branching. Comparing the two processes, we find that spin-correlations stay large for the $q$-initiated process in a wide region of phase-space. In fact, they only vanish when the two splittings are widely asymmetric, i.e., $z_{1,2}\to 0,1$. For the $g$-initiated channel the spin modulation falls off much faster than in the quark case and becomes negligible when $z_1\gtrsim0.8$. The second row of Fig.~\ref{fig:a_vac} shows results for a second splitting whose opening angle is two-times smaller than the associated dead-cone angle and thus mass effects are expected to play a role, see Eq.~\eqref{eq:adef-vac}. Indeed, we observe for both channels a significant depletion of spin-correlations by almost a factor 10.

\section{Collinear spin correlations in a QCD medium}\label{sec:in_medium_spin_corr}

\subsection{Heuristic discussion}



 
Let us now consider the same process $a \rightarrow b\,g \rightarrow b\,c\,d$, in the strongly ordered \textit{quasi}-collinear limit, inside a QCD medium. To simplify the calculation we take the $g\to c\,d$ splitting to be outside of the medium. Kinematically, this would correspond to a collinear splitting whose formation time is parametrically larger than the medium length, $L$.  
We describe the medium as a stochastic, coloured background field. At the amplitude level, each medium insertion yields a factor
\begin{equation}\label{eq:A_interaction}
i g t^{a}\gamma^{\mu} \cA^{a,\,\mu}(p^\mu)\,,
\end{equation}
where $g$ is the strong coupling, $a$ a colour index and $\cA$ the 
classical background field. Every parton in the $a \rightarrow b\,g \rightarrow b\,c\,d$ splitting will interact 
with such field and thus the amplitude squared for the process is expected to be modified with respect to vacuum. Before
carrying out the actual calculation of the in-medium amplitude, we discuss the analog of Eq.~\eqref{eq:real_imag_F_M2} for the medium case.

We consider partons with energies parametrically larger than any medium (e.g screening mass $m_D$) or transverse (relative to jet axis) scales. This means that parton-medium interactions do not significantly change the parton's virtuality, such that one can consider the intermediate gluon to stay nearly on-shell during its in-medium propagation. This hierarchy of scales further implies that we can 
neglect spin-flip terms so that $\bar u (p, \lambda)\gamma^\mu u(p, \lambda') \approx 2p^\mu\delta^{\lambda, \lambda'}$. Hence, one can assume the form of the amplitude in Eq.~\eqref{eq:general_factorised_M} to hold 
\begin{equation}\label{eq:M2_decomp_medium}
	\int_{\p_a}|\cM^{\rm med}_{a\to bg\to bcd}|^2 \propto \sum_{\lambda_g, \lambda_g'} F_{\rm med}^{\lambda_g \lambda_g'}(\q_1, \q_2)e^{i(\lambda_g-\lambda_g')\psi_{12}}\,,
\end{equation}
where we omitted the dependence on the remaining relevant variables, e.g. on $z_i$. The fact that in Eq.~\eqref{eq:M2_decomp_medium} the total transverse momentum dependence is integrated out is valid only for the case of transversely homogeneous media with instantaneous time correlations, see e.g.~\cite{Isaksen:2023nlr}. Unlike the vacuum case, $F_{\rm med}^{\lambda_g \lambda_g'}(\q_1, \q_2)$ now depends not only on the moduli of the $\q_i$ vectors but also on their orientations. This means, in particular, that one should expect medium modifications to introduce a non-trivial dependence on the angle between planes $\psi_{12}$. Since we take the second splitting to be outside the medium, the $\q_2$ dependence is given trivially by the vacuum expression for the second splitting
\begin{equation}
    F_{\rm med}^{\lambda_g \lambda_g'}(\q_1, \q_2)\underset{\substack{g\to cd \\ \text{in vacuum}}}{\rightarrow} F_{\rm med}^{\lambda_g \lambda_g'}(\q_1,|\q_2|)\,.
\end{equation}
We can now decompose $F_{\rm med}^{\lambda_g \lambda_g'}$ by projecting into the two possible helicity state combinations akin to the vacuum case, $F_{\rm med}^{\lambda_g \lambda_g'} = F_{\rm med,1}^{\lambda_g}\delta^{\lambda_g\lambda_g'}+F_{\rm med,2}^{\lambda_g}\delta^{\lambda_g, -\lambda_g'}$, and write the parametric dependence of the squared matrix element as
\begin{align}\label{eq:real_imag_F_M2_medium}
	\int_{\p_a}|\cM^{\rm med}_{a\to bg\to bcd}|^2 \propto  \sum_{\lambda_g}&\left[\Re F_{{\rm med},1}^{\lambda_g}(\q_{1},|\q_{2}|)+ \Re F_{{\rm med},2}^{\lambda_g}(\q_{1},|\q_{2}|)\cos(2\psi_{12}) \right.\nn
    &\left. - \lambda_g\,\Im  \,F_{{\rm med},2}^{\lambda_g}(\q_{1},|\q_{2}|)\sin(2\psi_{12})\right]\,.
\end{align}
We thus observe two major, qualitative modifications on the squared matrix-element with respect to vacuum: (i) it now depends on the orientation of $\q_1$, or equivalently on the value of $\phi_1$, and (ii) a potential non-zero $\sin(2\psi_{12})$ term if $F_{\rm med,2}^{\lambda_g}$ has a polarisation dependence. On top of these two effects, the dependence of $F_{\rm med,i}^{\lambda_g}$ on $z_1$ and $|\q_1|$ can also be modified with respect to vacuum.  
As argued in App.~\ref{app:vertex}, effects (i) and (ii) are only realized for a medium model that is not rotationally invariant, e.g., a medium with some preferred axis, see Eq.~\eqref{eq:squared_vertex_polarisation}. 
Thus, if one further considers an isotropic medium, the dependence on $\phi_{1}$ and the $\sin(2\psi_{12})$-term vanish and one is left with 
\begin{equation}\label{eq:real_imag_F_M2_medium_iso}
	\int_{\p_a}|\cM^{\rm med}_{a\to bg\to bcd}|^2 \underset{\substack{\text{isotropic} \\ \text{medium}}}{\to} \propto  \sum_{\lambda_g}\left[\Re F_{{\rm med},1}^{\lambda_g}(|\q_{1}|,|\q_{2}|)+ \Re F_{{\rm med},2}^{\lambda_g}(|\q_{1}|,|\q_{2}|)\cos(2\psi_{12})\right]\,.
\end{equation}
To summarise: in an anisotropic medium, we expect a $\sin(2\psi_{12})$-term together with a non-trivial dependence of the cross-section on $\phi_1$. In turn, for an isotropic medium, the azimuthal modulation is proportional to $\cos(2\psi_{12})$ as in vacuum, with both an overall modification and a change in the relative magnitude of the $\psi_{12}$-independent and $\cos(2\psi_{12})$-terms that will depend on both the phase-space point of the first splitting and the medium properties. We calculate these functional dependencies explicitly in the following sections.  

\subsection{Amplitude in the BDMPS-Z formalism}



Following the previous heuristic discussion, we now compute the medium modifications to the process $a\rightarrow b\,g \rightarrow b\,c\,d$, under the approximation that only the $a\rightarrow b\,g$ splitting occurs inside the medium and taking the medium to be static, homogeneous and of fixed length. 

The problem of a highly energetic particle propagating in a dense QCD medium can be tackled within the perturbative BDMPS-Z approach~\cite{Zakharov:1996fv,Baier:1996kr}\footnote{See e.g.~\cite{Mehtar-Tani:2013pia,Blaizot:2015lma,Apolinario:2022vzg, Mehtar-Tani:2025rty} for recent reviews.}, where one resums multiple scatterings with the medium at leading eikonal order, i.e., keeping the leading terms in powers of any transverse scale over the energy of the particle, while keeping sub-eikonal phases which are enhanced by the medium length and which are necessary to reproduce the Landau-Pomerachuk-Migdal effect~\cite{Wang:1994fx,Zakharov:1996fv}. Thus, in a given diagram, each line in colour representation $R$ is described by an effective non-relativistic quantum mechanical propagator, which in the light-cone gauge takes the form
\begin{align}\label{eq:def_in_medium_propagator}
	G_R^{ij}(t_2,\x_2;t_1,\x_1|p^+) = \int^{\boldsymbol{r}(t_2) = \boldsymbol{x_2}}_{\boldsymbol{r}(t_1)= \boldsymbol{x_1}}\mathcal{D}\boldsymbol{r}(t)\Exp{i\frac{p^+}{2}\int_{t_1}^{t_2}dt \, \dot{\boldsymbol{r}}^2}U_R^{ij}(t_1,t_2,\boldsymbol{r}(t)) \, ,
\end{align}
with the endpoints denoting the starting and final transverse positions of the particle's evolution in light-cone time, $\boldsymbol{r}(t=x^+)$, at fixed light-cone energy $p^+$. In the case of massive quarks one needs to account for mass dependent terms appearing in the propagator poles that lead to an additional sub-eikonal phase in Eq.~\eqref{eq:def_in_medium_propagator} that depends on the quark mass $m$
\begin{equation}\label{eq:massive_phase}
   G_F^{ij}(t_2,\x_2;t_1,\x_1|p^+) \rightarrow G_F^{ij}(t_2,\x_2;t_1,\x_1|p^+)\,  \exp[-i\frac{m^2}{2p^+}(t_2-t_1)]\, .
\end{equation}
As mentioned in the previous section, we assume $p^+ \gg |\p|,\,m,\,m_D$ to be the dominant scale of the problem, such that the relevant dynamics takes place in the plane transverse to the jet axis.\footnote{Light-cone energy $p^+$ is conserved throughout the particle's in-medium propagation.} Thus, the massive phase and the diffusive transverse motion dictated by the path integral are the leading sub-eikonal corrections. Further, the colour state of the parton is continuously rotated by successive scatterings with the medium and this is described by the path-ordered Wilson line
\begin{equation}\label{eq:def_wilson_line}
    U_R^{ij}(x^+,y^+,\boldsymbol{r}) = \mathcal{P}\Exp{ig\int_{x^+}^{y^+}ds^+\, \cA_a^{-}(s^+, \boldsymbol{r}(s^+))T^a_{ij}}\, ,
\end{equation}
where $T^a_{ij} = t^a_{ij}$ if $R=F$ (quark) and $T^a_{ij} = -if^{aij}$ if $R=A$ (gluon). Here $\mathcal{A}$ denotes an external classical field describing the underlying QCD background, as already introduced in Eq.~\eqref{eq:A_interaction}. To calculate any amplitude at leading eikonal accuracy, the propagators of the form of Eq.~\eqref{eq:def_in_medium_propagator} can be directly convoluted with the standard scalar QCD vertices (see App.~\ref{app:vertex}) such that, at tree level, one can use the same diagrams as in perturbation theory in the absence of a background medium, replacing the scalar particle propagators by in-medium propagators (cf. e.g.~\cite{Blaizot:2012fh,Isaksen:2023nlr,Barata:2024bqp,Mehtar-Tani:2025rty}).

The classical field $\cA$ is induced by a stochastic distribution of colour sources of the background medium. As such, the computation of any in-medium process should involve an average over possible configurations of colour sources. It is standard to assume that these configurations follow a Gaussian functional form, such that the only non-trivial average of $\cA$ fields is the two-point average given by~\cite{McLerran:1993ni,Gyulassy:1993hr}
\begin{equation}\label{eq:A_correlator} 
    \langle \mathcal{A}_a^-(x^+,x^-,\boldsymbol{x})\mathcal{A}_b^-(y^+,y^-,\boldsymbol{y})\rangle = \delta_{ab}n(x^+,\x)\delta(x^+-y^+)\gamma(\x-\y)\, ,
\end{equation}
which is local in colour and light-cone time. The form of field correlations is then set by $\gamma(\x)$, which is directly related to the in-medium elastic scattering rate, and by the medium density $n(x^+,\x)$, which for a static, fixed length and homogeneous medium is given by $n(x^+,\x) = n\,\Theta(x^+<L^+)$, where $L= L^+/\sqrt{2}$ is usually of the order of a few fm. Hence, any in-medium cross-section involves an average of squared amplitudes according to Eq.~\eqref{eq:A_correlator}.

Equipped with effective Feynman rules and a medium model we can now write the expression for the amplitude describing the $a\rightarrow b\,g \rightarrow b\,c\,d$ process. Consider particle $a$ to be produced at fixed light-cone time $t_0 \equiv 0$ from an initial amplitude $\cM_0$, with fixed light-cone energy $p_a^+$ and fixed, arbitrary polarisation state. In the mixed space of light-cone time and transverse momentum, the amplitude for the process $a\rightarrow b\,g\rightarrow b \,c \,d$, where the splitting $g\rightarrow c\,d$ is imposed to happen outside of the medium, can be written as
\begin{align}\label{eq:med_M}
	 i\cM &= \sum_{\lambda_g}\frac{e^{i\frac{\p_b^2}{2p_b^+}t_{\infty}}}{4p_a^+p_g^+}[igT_{ij}^l V_{g\rightarrow cd}^{\lambda_g \lambda_c \lambda_d}(z_2,\q_{2})] \int_0^{t_{\infty}} dt_1\,e^{-\varepsilon t_1}e^{i\frac{z_1^2m_1^2}{2\omega_a}t_1}\int_{t_1}^{t_{\infty}} dt_2\,e^{-\varepsilon t_2}\Theta(t_2 > L)\, \nn
    &\times e^{i\left(p_c^- + p_d^-\right)t_2} \int_{\k_b,\k_g}G^{lm}_{A}(t_2,\p_g;t_1,\k_g|p_g^+) G_{R_a}^{kn}(t_{\infty},\p_b;t_1,\k_b|p_b^+) [igT^m_{np}V_{a\rightarrow gb}^{\lambda_a \lambda_g \lambda_b}(z_1,\bkappa)]\, \nn
    &\times \int_{\p_0} G_{R_a}^{pq}(t_1,\k_0;0,\p_0|p_a^+)\cM_0^{q}(\vec p_0) \, ,
\end{align}
where $T^l_{jk} = t^l_{jk}\,(-if^{ljk})$ in the fundamental (adjoint) representations and $R_a = F \,(A)$ if the initial particle is a quark (gluon). Most kinematic variables have been defined in Sec.~\ref{sec:vac_spin_corr}, while the remaining ones are  $p_c^- = (\p_c^2 + m_2^2)/(2p_c^+)$, $p_d^- = (\p_d^2 + m_2^2)/(2p_d^+)$, $\p_g = \p_c+\p_d$ and the intermediate momenta given by $\k_0 = \k_b + \k_g$, $\vec p_0 = (p_a^+,\p_0)$ and $\bkappa = (1-z_1)\k_g - z_1\k_b$. In our approximation, the light-cone energy is conserved, $p_0^+ = p_a^+$, but the total transverse momentum is not, $\p_0 \neq \p_a = \p_b+\p_c+\p_d$. Further, note that only the mass phases attached to the light-cone times of the vertices survive, since the final state ones cancel after being put on-shell. Finally, the $\varepsilon$-prescription ($\varepsilon\rightarrow 0$) assures that medium interactions are adiabatically turned off~\cite{Wiedemann:2000za}, such that one recovers the correct vacuum limit when taking $t_\infty \rightarrow +\infty$. 

The integration variables $t_1$ and $t_2$ in Eq.\eqref{eq:med_M} correspond to the light-cone times of the first and second splittings, respectively. As discussed, we take the second splitting to be outside the medium and so impose the condition $\Theta(t_2>L)$ in Eq.~\eqref{eq:med_M}. This restriction on the $t_2$ integral naturally ignores contributions to the medium modification which would only be included in the full $1\rightarrow 3$ calculation inside the medium, which is considerably more involved. We thus expect our result to capture the correct qualitative behaviour in the case where there is a strong hierarchy on the formation time of the second splitting ($\tau_2$) with respect to the first splitting's ($\tau_1$) and the medium length, i.e.,
\begin{align}\label{eq:form_time_ordering}
    \tau_2 \gg L,\,\tau_1 \,, \quad \text{with}\quad \tau_1 \equiv 2p_a^+/(Q_a^2 -m_1^2)\,,\,\, \tau_2 \equiv 2p_g^+/Q_g^2 \,.
\end{align}
Additionally, note that this restriction on the $t_2$ integration 
introduces spurious terms 
when trying to recover the vacuum limit at the amplitude level. Indeed, replacing all propagators by their vacuum counterpart, i.e.,
\begin{equation}
    G_R^{ij} \rightarrow 
	\delta^{ij}G_0(t_2,\q;t_1,\p|p^+) = \delta^{ij}\,e^{-i\frac{\p^2}{2p^+}(t_2-t_1)}(2\pi)^2\delta^2(\q-\p)\, ,
\end{equation}
leads to the following result
\begin{equation}\label{eq:M_vac_limit}
    i\cM_{a\to bcd} \underset{\substack{\text{vacuum} \\ \text{limit}}}{\to} i\cM_{a\rightarrow bcd}^{\rm vac}\left[\frac{\tau_2}{\tau_2-\tau_1}e^{iL/\tau_2} - \left(\frac{\tau_2}{\tau_2-\tau_1} - 1\right)e^{iL/\tau_1} \right]\,,
\end{equation}
where $\cM^{\rm vac}_{a\rightarrow bcd}$ is the vacuum amplitude calculated in Eq.~\eqref{eq:vac_M}. Hence, in order to recover the exact vacuum limit in Eq.~\eqref{eq:M_vac_limit} we conclude that we need to impose the same formation time hierarchy as in Eq.~\eqref{eq:form_time_ordering}.
As long as one does not have $z_1\rightarrow 1$ (intermediate gluon taking all the energy on the first branching), this strong time-ordering is equivalent to the strong angular-ordering imposed in Eq.~\eqref{eq:ang_ordering}.

Finally, the light-cone time locality of the two-point average in Eq.~\eqref{eq:A_correlator} makes it convenient to separate the amplitude into two contributions: $\cM^{\rm in}$, where the first splitting is inside the medium, i.e., $t_1 \leq L$, and $\cM^{\rm out}$, where $t_1 > L$. The former contribution reads
\begin{align}\label{eq:M_in}
	 i\cM^{\rm in} &= \sum_{\lambda_g}\frac{e^{i\left(Q_a^2 -m_1^2 + \p_a^2\right)L/(2p_a^+)}}{2p_a^+}\left[\frac{i}{Q_{g}^2} igT_{ij}^l V_{g\rightarrow cd}^{\lambda_g \lambda_c \lambda_d}(z_2,\q_2)\right]\int_0^L dt_1 e^{-i\frac{z_1^2m_1^2}{2\omega_a}(L-t_1)} \, \nn
	& \times \int_{\k_b,\k_g,\p_0} G^{lm}_{A}(L,\p_g;t_1,\k_g|p_g^+) G^{kn}_{R_a}(L,\p_b;t_1,\k_b|p_b^+)[igT^m_{np}V_{a\rightarrow gb}^{\lambda_a \lambda_g \lambda_b}(z_1,\bkappa)]\, \nn
    & \times \int_{\p_0} G^{pq}_{R_a}(t_1,\k_0;0,\p_0|p_0^+)\cM_0^{q}(\vec p_0)\, ,
\end{align}
while the latter is given by
\begin{align}\label{eq:M_out}
	 i\cM^{\rm out}& = \sum_{\lambda_g}e^{i\left(Q_a^2 -m_1^2 + \p_a^2\right)L/(2p_a^+)}\left[\frac{i}{Q_{g}^2}igT_{ij}^lV_{g\rightarrow cd}^{\lambda_g \lambda_c \lambda_d}(z_2,\q_{2})\right]\left[\frac{i}{Q_a^2-m_1^2}  igT^l_{kp}V_{a\rightarrow gb}^{\lambda_a \lambda_b\lambda_g}(z_1,\q_{1})\right]\nn
    & \times \int_{\p_0} \,G^{pq}_{R_a}(L,\p_a;0,\p_0|p_a^+)\cM_0^{q}(\vec p_0)\, .
\end{align}

\subsection{Cross-section computation}\label{subsec:medium_xsec}
The fully differential in-medium cross-section for the process $a\rightarrow b\,g\rightarrow b\,c\,d$ is computed by first squaring the amplitude in Eq.~\eqref{eq:med_M}.
Following Eqs.~\eqref{eq:M_in},\eqref{eq:M_out}, we separate the calculation of the cross-section into time intervals according to
\begin{equation}\label{eq:M_region_separation}
	\cM\cM^{\dagger} = \cM^{\rm in}\cM^{\rm in, \dagger} + 2\Re\,\left(\cM^{\rm in}\cM^{\rm out, \dagger}\right)+\cM^{\rm out}\cM^{\rm out, \dagger}\,.
\end{equation}
The matrix-element squared then needs to be averaged over quantum numbers and over medium configurations ($\langle \,\cdot\, \rangle_\cA$) according to Eq.~\eqref{eq:A_correlator}. Combining this matrix-element with the corresponding phase-space element in App.~\ref{app:kinematics_phase_space}, leads to the fully differential cross-section 
\begin{equation}\label{eq:medium_xsec}
	\frac{16(2\pi)^6}{z_1^3(1-z_1)z_2(1-z_2)\theta_{1}\theta_{2}}\frac{d\sigma}{d^6\Omega_{\rm PS}} = \frac{1}{2}\int\, d^3\Omega_a (p_a^+)^4\sum_{\rho_i, \lambda_i}\left\langle \cM\cM^{\dagger} \right\rangle_{\cA}\,,
\end{equation}
where the sum runs over colours ($\rho_i$) and spin states ($\lambda_i$), and the factor $1/2$ accounts for the average over the initial parton spin state.

Naturally, the term $d\sigma^{\rm out-out}$ simply reduces to the vacuum result given by Eq.~\eqref{eq:xsection-vac}, since both splittings happen outside the medium and the broadening of the initial parton is integrated out due to the integration over total transverse momentum $\p_a$.
To compute the remaining terms, we note that from the medium-averaging point of view, only the first splitting matters, since the second occurs outside of the medium. We also adopt the large-$N_c$ limit throughout the calculation. Then, Eq.~\eqref{eq:A_correlator} dictates that: (i) medium correlations are local in time and colour, and (ii) the medium is transversely homogeneous ($\gamma(\x,\y) = \gamma(\x-\y))$. Point (i) implies that the amplitude $\mathcal{M}$ and its complex conjugate amplitude $\mathcal{M}^\dagger$ form a colour singlet state at all times, while points (i) and (ii) imply that the total transverse momentum must be the same in $\mathcal{M}$ and $\mathcal{M}^\dagger$ at all times (cf.~\cite{Blaizot:2012fh, Isaksen:2023nlr}). 
Finally, by integrating over total momentum information (integral in $d^3\Omega_a$), the broadening of the initial parton $a$ is integrated out and the squared initial production amplitude can be factored out. Thus, following a similar notation to~\cite{Isaksen:2023nlr}, the $d\sigma^{\rm in-in}$ and $d\sigma^{\rm in-out}$ cross-section contributions read,
\begin{subequations}
\begin{align}
	& (2\pi)^2\frac{d\sigma^{\rm in-in}}{\sigma_0\,d^6\Omega_{\rm PS}} = \left(\frac{\alpha_s}{\pi}\right)^2\frac{C_1\theta_{1}}{8}\frac{C_2\omega_g^2 \theta_{2}P^M_{g\rightarrow cd}(z_2,\theta_{2})}{2z_2(1-z_2)Q_g^2}\nn
    & \times 2\Re \int_0^L dt_1 \int_{t_1}^L d\bar t_1 e^{-i\frac{z_1^2m_1^2}{2\omega_a}(\bar t_1 - t_1)}\int_{\bkappa,\bar \bkappa, \l} \cQ_{a\rightarrow bg}(L,\q_{1}, \q_{1}; \bar t_1, \bar \bkappa, \l)\cK_{a\rightarrow bg}(\bar t_1, \l; t_1, \bkappa) \nn
    & \times \Big[z_1^3m_1^2 + P_{a\rightarrow gb}(z_1)\left(\bkappa\cdot \bar\bkappa\right)+ S_{g\to cd}\frac{4z_2(1-z_2)(1-z_1)}{z_1(\q_{2}^2P_{g\rightarrow cd}(z_2)+m_2^2)}\Big((\q_{2}\cdot \bkappa)(\q_{2}\cdot \bar\bkappa)
    \nn
    & -(\q_{2}\times \bkappa)_z(\q_{2}\times \bar\bkappa)_z\Big)\Big] \, ,
    \label{eq:M_in_in2}
    \\
    & (2\pi)^2\frac{d\sigma^{\rm in-out}}{\sigma_0\,d^6\Omega_{\rm PS}} = -\left(\frac{\alpha_s}{\pi}\right)^2\frac{C_1\omega_a^2 \theta_{1}}{2z_1(1-z_1)(Q_a^2-m_1^2)}\frac{C_2\omega_g^2 \theta_{2}P^M_{g\rightarrow cd}(z_2,\theta_{2})}{2z_2(1-z_2)Q_g^2}\frac{1}{2\omega_a}\nn
    & \times 2\Re \,i\int_0^L dt_1 e^{-i\frac{z_1^2m_1^2}{2\omega_a}(L-t_1)}\int_{\bkappa} \cK_{a\rightarrow bg}(L, \q_1; t_1, \bkappa)\Big[z_1^3m_1^2 + P_{a\rightarrow gb}(z_1)\left(\bkappa\cdot \q_1\right)\nn
    & +S_{g\to cd}\frac{4z_2(1-z_2)(1-z_1)}{z_1(\q_{2}^2P_{g\rightarrow cd}(z_2)+m_2^2)} \Big((\q_{2}\cdot \bkappa)(\q_{2}\cdot \q_1)-(\q_{2}\times \bkappa)_z(\q_{2}\times \q_1)_z\Big)\Big] \, ,
     \label{eq:M_in_out2}
    \end{align}
\end{subequations}
where $(\q_2\times \bkappa)_z = q_2^x\kappa^y-q_2^y\kappa^x$ and, as in the vacuum case, $S_{g\rightarrow gg} = +1$ and $S_{g\rightarrow q\bar q} = -1$. 

Let us discuss some of the ingredients that enter into Eqs.~\eqref{eq:M_in_in2} and \eqref{eq:M_in_out2}. First, we observe that the vacuum splitting function for the second branching, $P^M_{g\to cd}$, appears. The terms within brackets inside the $\bkappa$-integral correspond to the polarisation sum of the vertex product squared (see Eq.~\eqref{eq:AAdagger_all_channels}). To factor out $\sigma_0$ (defined in Eq.~\eqref{eq:sigma_0_def}) we have assumed that $\cM_0$ is sharply peaked around some value of energy $p_a^+$. The objects $\cK$ and $\cQ$, defined in App.~\ref{app:npoint_avs}, involve only two-body (dipole) and four-body (quadrupole) medium averages, which is to be expected in the large-$N_c$ limit (cf.~\cite{Dominguez:2011wm,Dominguez:2012ad}). The latter can be written as a sum of a term with only dipoles ("factorizable") and a term containing quadrupoles ("non-factorizable")~\cite{Blaizot:2012fh,Apolinario:2014csa,Dominguez:2019ges,Isaksen:2023nlr}. In what follows, we shall neglect the "non-factorizable" piece\footnote{This approximation was tested quantitatively in the context of $\gamma \rightarrow q\bar q$~\cite{Isaksen:2023nlr}. It was concluded that the relative magnitude of the non-factorizable piece can get large when considering sufficiently large splitting angles $\theta \gtrsim 0.4$ and finite energy fractions $z \gtrsim 0.1$. Thus we expect the large-$N_c$ result with only the factorizable piece to correctly capture the soft and collinear regime of the spectrum, while reasonably describing the main qualitative features of medium modifications outside of this region. Note that although the colour structure of $\gamma \rightarrow q\bar q$ is much simpler than that of pure QCD processes, this process still provides an indication of whether neglecting the non-factorizable contribution is reasonable. This is because at large-$N_c$ the quadrupole arising from the 4-body average is common to all processes $\gamma \rightarrow q\bar q$, $q \rightarrow gq$ and $g \rightarrow gg$ (cf. e.g.~\cite{Blaizot:2012fh,Apolinario:2014csa,Dominguez:2019ges}).}, which amounts to taking
\begin{align}\label{eq:Q_approx}
	\cQ_{a\rightarrow bg}(L,\q_1, \q_1; \bar t_1, \bar \bkappa, \l) \approx \delta^2(\l-\bar\bkappa) \int_\v e^{-i\v\cdot(\q_1-\bar\bkappa)}\cP_{a\rightarrow gb}(L,\bar t_1, \v)\,,
\end{align}
where $\cP$ is a product of two dipole objects, describing the independent broadening of the intermediate gluon and parton $b$ (see Eq.~\eqref{eq:indep_broadening}). Under this approximation, all medium averages can be written as functions of products of dipole objects of the form
\begin{align}\label{eq:dipole-av}
	D(\bar t, t; \u) = \frac{1}{N_c}\left\langle\Tr\left(U_1^{\dagger}U_2\right)\right\rangle = \Exp{-\frac{C_F}{2}\int_{t}^{\bar t} ds\, n(s) \sigma(\u)}\,, \quad \u = \r_1-\r_2\, ,
\end{align}
where $U_i^{jk} = U_F^{jk}(\bar t, t; \r_i)$ as defined in Eq.~\eqref{eq:def_wilson_line} and the dipole cross-section is written in terms of the scattering rate potential entering Eq.~\eqref{eq:A_correlator} as $\sigma(\r) = 2g^2(\gamma(0)-\gamma(\r))$. 

Plugging Eq.~\eqref{eq:Q_approx} into Eqs.~\eqref{eq:M_in_in2},\eqref{eq:M_in_out2} and performing the Fourier transforms to write everything in coordinate space we obtain:
\begin{subequations}
\begin{align} 
	& (2\pi)^2\frac{d\sigma^{\rm in-in}}{\sigma_0\,d^6\Omega_{\rm PS}} = \left(\frac{\alpha_s}{\pi}\right)^2\frac{C_1\theta_{1}}{8}\frac{C_2\omega_g^2 \theta_{2}P^M_{g\rightarrow cd}(z_2,\theta_{2})}{2z_2(1-z_2)Q_g^2}\nn
    & \times 2\Re\int_0^L dt_1 \int_{t_1}^L d\bar t_1 e^{-i\frac{z_1^2m_1^2}{2\omega_a}\Delta t}\int_{\v}\Big[z_1^3m_1^2 + P_{a\rightarrow gb}(z_1)\left(\nabla_{\u}\cdot \nabla_{\bar\u}\right)-\nn
    & + S_{g\to cd}\frac{4z_2(1-z_2)(1-z_1)}{z_1(\q_{2}^2P_{g\rightarrow cd}(z_2)+m_2^2)}\left(\left(\q_{2}\cdot \nabla_{\u}\right)\left(\q_{2}\cdot \nabla_{\bar\u}\right)-\left(\q_{2}\times \nabla_{\u}\right)_z\left(\q_{2}\times \nabla_{\bar\u}\right)_z\right)\Big]\nn
	& \times e^{-i\v\cdot\q_{1}}\cP_{a\rightarrow gb}(L,\bar t_1, \v)\left.\cK_{a\rightarrow bg}(\bar t_1, \bar\u; t_1, \u)\right|_{\u = 0, \bar\u = \v} \, ,
    \label{eq:med_xsec_before_HO_inin}
    \\
	& (2\pi)^2\frac{d\sigma^{\rm in-out}}{\sigma_0\,d^6\Omega_{\rm PS}} = \left(\frac{\alpha_s}{\pi}\right)^2\frac{C_1\omega_a^2 \theta_{1}}{2z_1(1-z_1)(Q_a^2-m_1^2)}\frac{C_2\omega_g^2 \theta_{2}P^M_{g\rightarrow cd}(z_2,\theta_{2})}{2z_2(1-z_2)Q_g^2}\frac{1}{2\omega_a}\nn
    & \times 2\Re\int_0^L dt_1e^{-i\frac{z_1^2m_1^2}{2\omega_a}(L-t_1)}\int_\v\Big[ z_1^3m_1^2 + P_{a\rightarrow gb}(z_1)\left(\nabla_{\u}\cdot \q_{1}\right)-\nn
    & + S_{g\to cd}\frac{4z_2(1-z_2)(1-z_1)}{z_1(\q_{2}^2 P_{g \rightarrow cd}(z_2) + m_2^2)}\left(\left(\q_{2}\cdot \nabla_{\u}\right)\left(\q_{2}\cdot \q_{1}\right)-\left(\q_{2}\times \nabla_{\u}\right)_z\left(\q_{2}\times \q_{1}\right)_z\right)\Big]\nn
	& \times e^{-i\v\cdot\q_{1}}\left.\cK_{a\rightarrow bg}(L, \v; t_1, \u)\right|_{\u = 0} \, .
    \label{eq:med_xsec_before_HO_inout}
    \end{align}
\end{subequations}

At this point we have to provide an explicit form for the dipole cross-section $\sigma(\r)$ entering the definition of the dipole medium average~\eqref{eq:dipole-av}. In the multiple soft scattering approximation one expands $\sigma(\r)$ in the small distance limit, keeping only its quadratic dependence. Under such approximation, the potential is fully characterized by the jet quenching parameter $\hat q$~\footnote{Throughout this paper, we use $\hat q$ in the adjoint representation.}, which in a medium with a momentum space anisotropy can take different values in the two directions orthogonal to the jet axis/initial particle's direction, i.e. $\hat q_x \neq \hat q_y$. This model for anisotropy has been implemented in this type of calculation in~\cite{Hauksson:2023tze, Barata:2024bqp,Barata:2025uxp}.\footnote{See e.g. Refs.~\cite{Barata:2022krd,Andres:2022ndd,Barata:2023qds,Barata:2023zqg, Kuzmin:2023hko,Kuzmin:2024smy} for different forms of including a medium with a non-trivial structure in a BDMPS-Z calculation.} For a static, finite length medium, the multiple soft scattering approximation takes the form
\begin{align}
    N_c \int_{t_1}^{t_2} dt \, n(s^+) \sigma(\boldsymbol{r}) & = N_c  (t_2-t_1) \,n\, \sigma(\boldsymbol{r}) \, \nn
    &= \frac{t_2-t_1}{2}\left(\hat{q}_x r_x^2 + \hat{q}_y r_y^2\right) + \mathcal{O}(\r^2\log(1/(m_D^2\r^2)) \, ,
\end{align}
where we have written the components of the transverse vector as $\r = (r_x,r_y)$ and $m_D^{-1}$ sets a screening length for interactions. The broadening kernel can then be written as
\begin{equation}\label{eq:P_after_HO}
	\cP_{a\rightarrow gb}(L,\bar t_1, \v) = \Exp{-\frac{1}{8}f_{a\rightarrow bg}(z_1)(L-\bar t_1)(\hat q_x v_x^2 + \hat q_y v_y^2)}\, .
\end{equation}
The splitting kernel $\cK_{a\rightarrow bg}(\bar t_1, \u_2; t_1, \u_1)$ has a well known solution in the literature (see e.g.~\cite{Kleinert:2004ev, Apolinario:2014csa})
\begin{align}\label{eq:K_after_HO}
	& \cK_{a\rightarrow bg}(\bar t_1, \u_2; t_1, \u_1) =  \sqrt{\frac{\omega_a\Omega_x}{2\pi i \sin{\Omega_x\Delta t}}}\Exp{i \frac{\omega_a\Omega_x}{2 \sin{\Omega_x\Delta t}} \left[ \left(u_{1x}^2+u_{2x}^2\right)\cos{\Omega_x\Delta t} - 2u_{1x}u_{2x}\right]} \nn
    & \times \sqrt{\frac{\omega_a\Omega_y}{2\pi i \sin{\Omega_y\Delta t}}} \Exp{i \frac{\omega_a\Omega_y}{2 \sin{\Omega_y\Delta t}} \left[\left(u_{1y}^2+u_{2y}^2\right)\cos{\Omega_y\Delta t} - 2u_{1y}u_{2y}\right]} \, ,
\end{align}
with 
\begin{equation}\label{eq:omega}
    \Omega_i = \left(\frac{1-i}{\sqrt{2}}\right)\sqrt{\frac{\hat{q}_i h_{a\rightarrow bg}(z_1)}{4\omega_a}},  \qquad \Delta t = \bar t_1 - t_1 \,.
\end{equation}
The process dependent functions $f(z_1)$ and $h(z_1)$ read
\begin{subequations}\label{eq:process_dep_hf}
\begin{align}
    \begin{split}
    & f_{q\rightarrow g q}(z_1) = 2(1-z_1)^2 + z_1^2\,,\quad h_{q\rightarrow g q}(z_1)= 1 + (1-z_1)^2\, .
    \end{split}\\
    \begin{split}
    &f_{g\rightarrow g g}(z_1) = 2[(1-z_1)^2 + z_1^2]\,,\quad h_{g\rightarrow g g}(z_1)= 1 + z_1^2 + (1-z_1)^2\,.
    \end{split}
    \end{align}
\end{subequations}

Using the explicit form for the broadening and splitting kernels in Eqs.~\eqref{eq:P_after_HO} and~\eqref{eq:K_after_HO} and parametrizing the transverse momenta according to App.~\ref{app:kinematics_phase_space},  Eqs.~\eqref{eq:med_xsec_before_HO_inin} and ~\eqref{eq:med_xsec_before_HO_inout} can be calculated analytically modulus integrations over light-cone times. We can then write the full spectrum in the following form
\begin{align}\label{eq:xsec_med_full}
	 (2\pi)^2\frac{d\sigma^{\rm med}}{\sigma_0\,d^6\Omega_{\rm PS}}& = \left(\frac{\alpha_s}{\pi}\right)^2 C_1 C_2\frac{P^M_{a\rightarrow bg}(z_1,\theta_{1})}{\theta_1(1+\tilde\theta^2_{m,1})}\frac{ P^M_{g\rightarrow cd}(z_2,\theta_{2})}{\theta_2(1+\tilde\theta^2_{m,2})} (1+F_{\rm med}) 
     \nonumber \\
    & \times \left[1 + a_{\rm med} \cos(2\psi_{12} +\phi_{\rm med})\right]\, ,
\end{align}
where we observe (i) an overall modification of the spectrum encapsulated in the factor $1+F_{\rm med}$~\cite{Dominguez:2019ges,Isaksen:2023nlr}, (ii) a change in the modulation of the azimuth-dependent term ($a_{\rm med}$) and (iii) a phase-shift ($\phi_{\rm med}$). These 3 factors depend on a series of auxiliary functions that we proceed to write down explicitly:
\vspace{-0.5cm}
\begin{subequations}
\begin{align}
    & a_{\rm med} = a \frac{\sqrt{A_1^2+A_2^2}}{1+F_{\rm med}} \label{eq:amed-def}\,,\\
    & \tan\phi_{\rm med} = \frac{A_2}{A_1} 
    \label{eq:phimed-def}\,,
\end{align}
\end{subequations}
where $a$ corresponds to the amplitude of spin correlations in vacuum, see Eq.~\eqref{eq:adef-vac}. The auxiliary functions $A_1$, $A_2$ and $F_{\rm med}$ are given by 
\begin{subequations}\label{eq:Ai_coeff_aniso}
\begin{align}
	 F_{\rm med} &= \frac{1}{\omega_a P^M_{a\rightarrow bg}(z_1,\theta_{1})}\,2\Re\,i\,\Bigg\lbrace\int^L_{0} dt_1 e^{-i\frac{z_1^2m_1^2}{2\omega_a}(L-t_1)}e^{-i\frac{\q_{1}^2}{2\omega_a}\left(\frac{\cos^2\phi_1}{c_{2x}}+\frac{\sin^2\phi_1}{c_{2y}}\right)}\frac{\sqrt{c_{1x}}\sqrt{c_{1y}}}{\sqrt{c_{2x}}\sqrt{c_{2y}}} \nonumber \\
    &\times \Big[z_1^3m_1^2 + P_{a\rightarrow gb}(z_1)\left(\cD_+ + \cD_- \cos(2\phi_1)\right)\Big] - \left(\frac{Q_a^2-m_1^2}{2p_a^+}\right) \int_{0}^{L}dt_1 \int_{t_1}^{L}d\bar t_1\, e^{-i\frac{z_1^2m_1^2}{2\omega_a}\Delta t} 
    \nn
    &\times e^{i\frac{\q_{1}^2}{2\omega_a}\left(\frac{\cos^2\phi_1}{c_{3x}}+\frac{\sin^2\phi_1}{c_{3y}}\right)}  \frac{\sqrt{c_{1x}}\sqrt{c_{1y}}}{\sqrt{c_{3x}}\sqrt{c_{3y}}}\Big[z_1^3m_1^2 + P_{a\rightarrow gb}(z_1)\left(\cA_+ + \cB \cos(2\phi_1)\right) \Big]\Bigg\rbrace \,, \label{eq:fmed-def}\\ 
	 A_1 &= 1+\frac{\q_1^2P_{a\rightarrow gb}(z_1) + z_1^3m_1^2}{\q_1^2\omega_aP^M_{a\rightarrow bg}(z_1,\theta_{1})}\,2\Re\,i\,\Bigg\lbrace \int_{0}^L dt_1 e^{-i\frac{z_1^2m_1^2}{2\omega_a}(L-t_1)}e^{-i\frac{\q_{1}^2}{2\omega_a}\left(\frac{\cos^2\phi_1}{c_{2x}}+\frac{\sin^2\phi_1}{c_{2y}}\right)}\frac{\sqrt{c_{1x}}}{\sqrt{c_{2x}}}\nn
    & \times \frac{\sqrt{c_{1y}}}{\sqrt{c_{2y}}}\left[\cD_+ + \cD_- \cos(2\phi_1)\right]-\left(\frac{Q_a^2-m_1^2}{2p_a^+}\right)\int_{0}^L  d t_1\int_{t_1}^L d\bar t_1\,  e^{-i\frac{z_1^2m_1^2}{2\omega_a}\Delta t}\nn
    &\times e^{i\frac{\q_{1}^2}{2\omega_a}\left(\frac{\cos^2\phi_1}{c_{3x}}+\frac{\sin^2\phi_1}{c_{3y}}\right)}\frac{\sqrt {c_{1x}}\sqrt{c_{1y}}}{\sqrt{ c_{3x}}\sqrt{c_{3y}}} \left[\cC_+ + \cA_- \cos(2\phi_1) + \cC_-\cos(4\phi_1) \right]\Bigg\rbrace \,, \label{eq:a1-def}\\
	  A_2 &=-\frac{\q_1^2P_{a\rightarrow gb}(z_1) + z_1^3m_1^2}{\q_1^2\omega_aP^M_{a\rightarrow bg}(z_1,\theta_{1})}\,2\Re\,i\,\Bigg\lbrace \int_{0}^L dt_1\,e^{-i\frac{z_1^2m_1^2}{2\omega_a}(L-t_1)}e^{-i\frac{\q_{1}^2}{2\omega_a}\left(\frac{\cos^2\phi_1}{c_{2x}}+\frac{\sin^2\phi_1}{c_{2y}}\right)}\frac{\sqrt{c_{1x}}}{\sqrt{c_{2x}}}\,\nn
    &\times \frac{\sqrt{c_{1y}}}{\sqrt{c_{2y}}}\cD_-\sin(2\phi_1) -\left(\frac{Q_a^2-m_1^2}{2p_a^+}\right)\int_{0}^L dt_1\int_{t_1}^L d\bar t_1 \,e^{-i\frac{z_1^2m_1^2}{2\omega_a}\Delta t}e^{i\frac{\q_{1}^2}{2\omega_a}\left(\frac{\cos^2\phi_1}{c_{3x}}+\frac{\sin^2\phi_1}{c_{3y}}\right)} \nn
    &\times \frac{\sqrt{c_{1x}}\sqrt{c_{1y}}}{\sqrt{c_{3x}}\sqrt{c_{3y}}}\left[\cA_- \sin(2\phi_1) + \cC_-\sin(4\phi_1)\right]\Bigg\rbrace\,.
    \label{eq:a2-def}
    \end{align}
\end{subequations}
Finally, the complex $c$-functions, defined as
\begin{subequations} \label{eq:ci_definition_aniso}
\begin{align}
    &c_{1i} = \frac{\Omega_i}{2i\sin{\Omega_i\Delta t}}\,, 
    \label{eq:c1_definition_aniso}\\
    &c_{2i} = \frac{\Omega_i}{\tan{\Omega_i\Delta t}}\,,
   \label{eq:c2_definition_aniso}\\
     &c_{3i} = (L-\bar t_1) \left(-i\frac{\hat q_i f_{a\rightarrow bg}(z_1)}{4\omega_a}\right)-c_{2i}\,,
     \label{eq:c3_definition_aniso}
     \end{align}
   \end{subequations}
encode the medium anisotropy and enter into the remaining functions 
\begin{subequations}\label{eq:abcd-functions}
\begin{align}
\begin{split}
    &\cA_{\pm} = 2\omega_a\left(c_{1x} \pm c_{1y} + \frac{c_{1x}c_{2x}}{c_{3x}} \pm \frac{c_{1y}c_{2y}}{c_{3y}}\right) + i\q_{1}^2\left(\frac{c_{1x}c_{2x}}{c_{3x}^2} \pm \frac{c_{1y}c_{2y}}{c_{3y}^2}\right)\,,
    \end{split}\\
    \begin{split}
	&\cB = i\q_{1}^2\left(\frac{c_{1x}c_{2x}}{c_{3x}^2}-\frac{c_{1y}c_{2y}}{c_{3y}^2}\right)\,,
    \end{split}\\
    \begin{split}
	&\cC_{\pm} = \frac{i\q_{1}^2}{2}\left(\frac{c_{1x}}{c_{3x}}\pm\frac{c_{1y}}{c_{3y}}\right)\left(\frac{c_{2x}}{c_{3x}}\pm\frac{c_{2y}}{c_{3y}}\right)\,,
    \end{split}\\
     \begin{split}
	  & \cD_{\pm} = \q_{1}^2\left(\frac{c_{1x}}{c_{2x}}\pm\frac{c_{1y}}{c_{2y}}\right)\,.
       \end{split}
    \end{align}
\end{subequations}
Eqs.\eqref{eq:xsec_med_full}-\eqref{eq:abcd-functions} are the main result of this paper. A few comments are in order before presenting the numerical calculation of this cross-section. First, as a consequence of assuming only the $a\to b\,g$ splitting to be modified by the medium, the $A_1, A_2$ and $F_{\rm med}$ auxiliary functions solely depend on the first branching kinematics ($p_a^+,m_1,\theta_1,z_1,\phi_1$)  and on the medium parameters ($L$, $\hat q_x$, $\hat q_y$). Recovering the vacuum limit, effectively amounts to taking $1+F_{\rm med}\to 1, A_1\to 1, A_2\to 0$. The existence of odd parity harmonics in $\psi_{12}$ and $\phi_{1}$, i.e., $\sin(2\psi_{12})$, $\sin(2\phi_{1})$ and $\sin(4\phi_{1})$, is directly related, under our assumptions, to the medium anisotropy, as previously argued. Note that, in the context of the $1\rightarrow 2$ cross-section in anisotropic media, the 
odd parity harmonics in $\phi_1$ show up only when final-state spin dependence is taken into account~\cite{Barata:2024bqp}.
Regarding the direction of anisotropy, we observe that if one inverts it (i.e. $\hat q_x \leftrightarrow \hat q_y$), then the resulting $\phi_{1}$ distribution is the same as if one transforms it under $\phi_{1} \rightarrow \phi_{1}+\pi/2$. To see this, note that upon switching $\hat q_x \leftrightarrow \hat q_y$, the coefficients $\cA_-,\,\cB,\,\cD_-$ gain a minus sign, while the remaining stay unchanged. All three coefficients are attached to an even harmonic ($\cos(2\phi_1)$ and $\sin(2\phi_1)$), which gains a minus sign under $\phi_1\rightarrow \phi_1 + \pi/2$, canceling the sign from the coefficient. The $\phi_1$ dependent phase is trivially invariant under these two simultaneous transformations. 
\paragraph{Isotropic medium} Setting $\hat q_x = \hat q_y$ in Eqs.~\eqref{eq:xsec_med_full}-\eqref{eq:abcd-functions} leads to a vanishing phase-shift $\phi_{\rm med}=0$ together with $A_2=0$, i.e., $a_{\rm med}=a A_1/(1+F_{\rm med})$. The auxiliary functions reduce to
\begin{subequations}
\begin{align}
& F_{\rm med} \underset{\hat q_x=\hat q_y}{=}\frac{1}{\omega_a P^M_{a\rightarrow bg}(z_1,\theta_1)}\,2\Re\,i\,\Bigg\lbrace \int_{0}^L dt_1  e^{-i\frac{z_1^2m_1^2}{2\omega_a}(L-t_1)}e^{-i\frac{\q_1^2}{2\omega_ac_2}}\frac{c_1}{c_2}\left[z_1^3m_1^2 + P_{a\rightarrow gb}(z_1)\cD_+\right] \nonumber \\ 
&- \left(\frac{Q_a^2-m_1^2}{2p_a^+}\right)\int_{0}^L dt_1 \int_{t_1}^L d\bar t_1\, e^{-i\frac{z_1^2m_1^2}{2\omega_a}\Delta t}e^{i\frac{\q_1^2}{2\omega_ac_3}} \frac{c_1}{c_3}\left[z_1^3m_1^2 + P_{a\rightarrow gb}(z_1)\cA_+\right]\Bigg\rbrace \, ,
\label{eq:fmed-iso}
\\
	& A_1 \underset{\hat q_x=\hat q_y}{=} 1+\frac{\q_1^2 P_{a\rightarrow gb}(z_1) + z_1^3m_1^2}{\q_1^2\omega_aP^M_{a\rightarrow bg}(z_1,\theta_1)}\,2\Re\,i\,\Bigg\lbrace\int_{0}^L dt_1 e^{-i\frac{z_1^2m_1^2}{2\omega_a}(L-t_1)}e^{-i\frac{\q_1^2}{2\omega_ac_2}}\frac{c_1}{c_2}\cD_- \nonumber\\
    & -\left(\frac{Q_a^2-m_1^2}{2p_a^+}\right)\int_{0}^L dt_1 \int_{t_1}^L d\bar t_1\, e^{-i\frac{z_1^2m_1^2}{2\omega_a}\Delta t}e^{i\frac{\q_1^2}{2\omega_ac_3}} \frac{c_1}{c_3}\cC_+\Bigg\rbrace\, .
    \label{eq:A1-iso}
\end{align}
\end{subequations}

\section{Numerical results}
\label{sec:numerics}
We now numerically evaluate Eq.~\eqref{eq:xsec_med_full}, and study the impact of medium modifications on the spin correlations modulation's amplitude $a_{\rm med}$ and their phase shift $\phi_{\rm med}$, as well as the total cross-section as a function of the azimuthal angles between planes $\psi_{12}$. This is done for both an isotropic (Sec.~\ref{subsec:isotropic}) and an anisotropic medium (Sec.~\ref{subsec:anisotropic}) with a fixed medium length of $L=5$ fm. Similarly to the vacuum results in Sec.~\ref{subsec:vac_xsec}, we take the first splitting to be massless ($m_1 = 0$) and fix its energy to $E_a=100$ GeV. We emphasize that the second splitting's kinematics $(z_2,\theta_2)$ only affects the vacuum piece in Eq.~\eqref{eq:xsec_med_full}. Finally, as mentioned above, we do not provide numerical results for $F_{\rm med}$, since this has been addressed multiple times in the literature (see e.g.~\cite{Dominguez:2019ges,Isaksen:2023nlr}).
\begin{figure}
     \centering
         \includegraphics[width=0.49\textwidth]{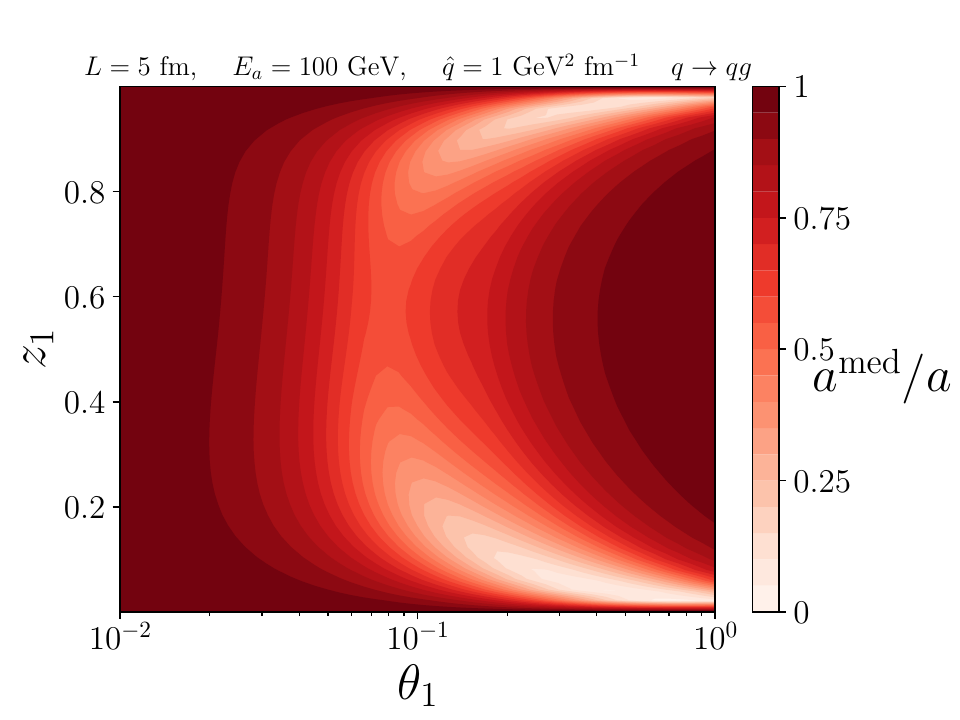}
         \includegraphics[width=0.49\textwidth]{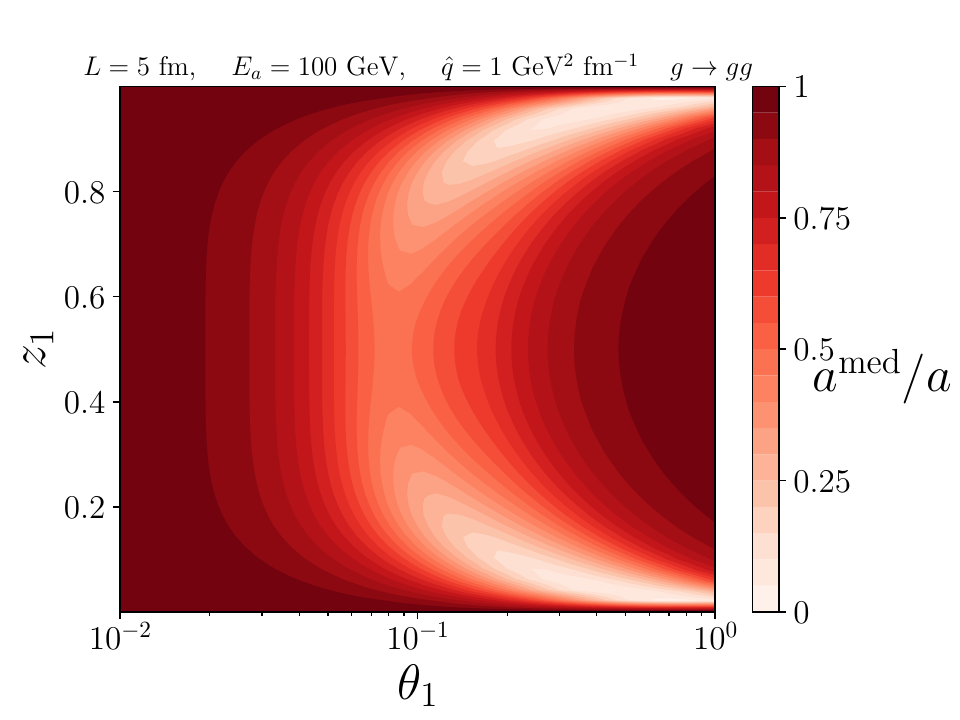}
         \\
         \includegraphics[width=0.49\textwidth]{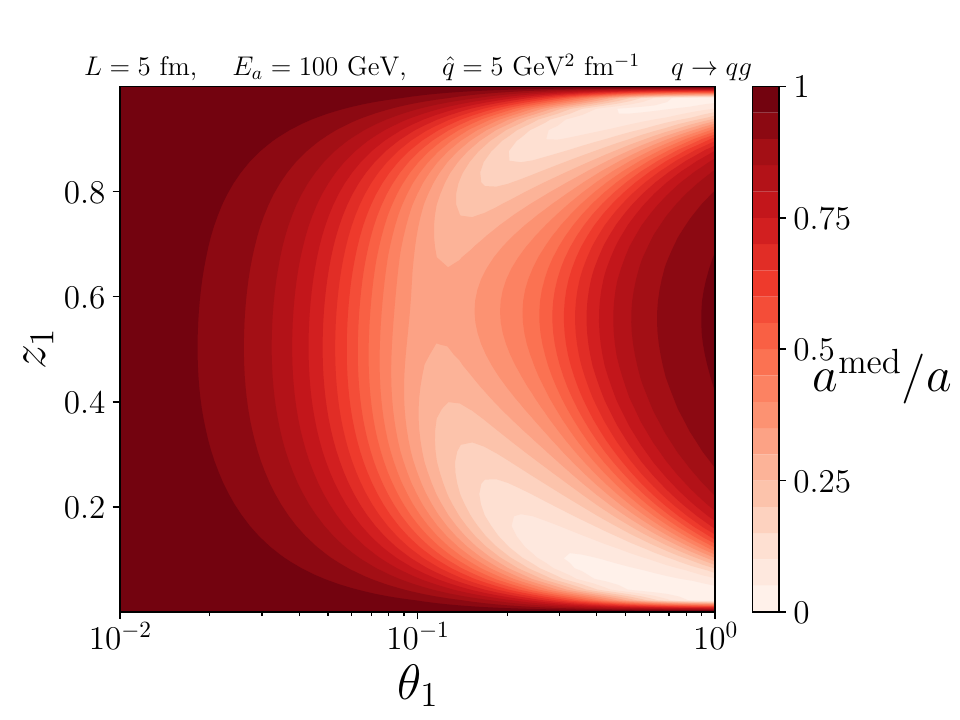}
         \includegraphics[width=0.49\textwidth]{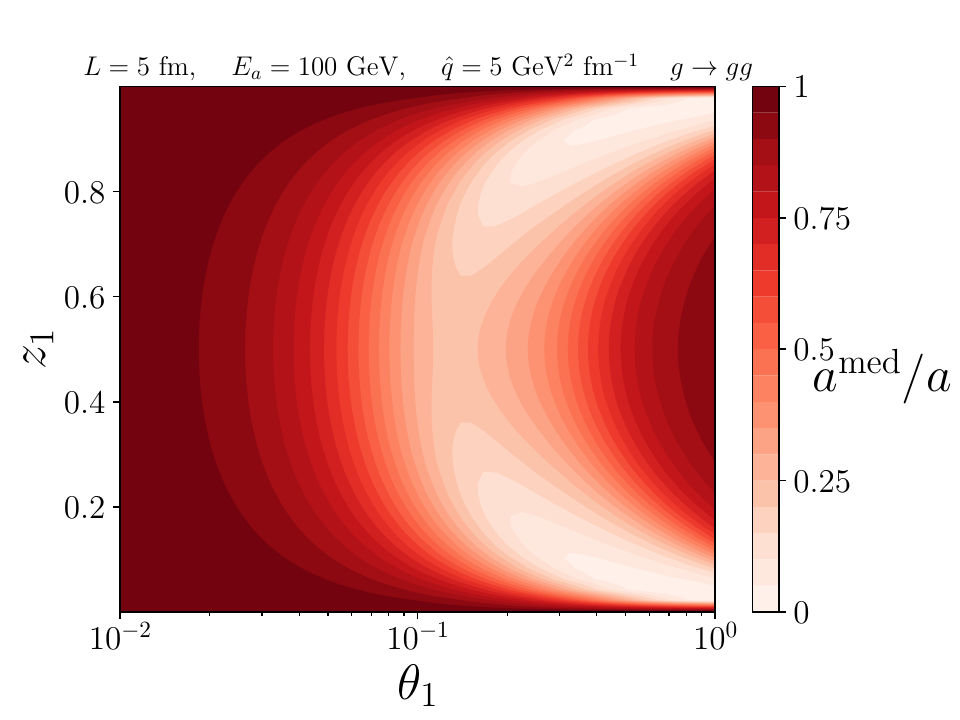}
     \caption{Ratio between the the spin-correlations modulation amplitude in the medium and in vacuum $a_{\rm med}/a$ as a function of the first splitting's kinematics $z_1$ and $\theta_1$. The upper panels correspond to the first splitting being $q\rightarrow qg$ and the lower ones to $g\rightarrow gg$. The medium is isotropic and we plot $\hat q= 1$ GeV$^2$/fm (upper panels) and $5$ GeV$^2$/fm (lower panels).}
     \label{fig:A2_over_A1_iso}
\end{figure}

\subsection{Isotropic medium}\label{subsec:isotropic}

In Fig.~\ref{fig:A2_over_A1_iso}, we show the medium-induced modification on the spin-correlations signal, $a_{\rm med}/a$, as a function of the first splitting's kinematics $(z_1,\theta_1)$ and for two values of $\hat q$: $\hat q = 1$ GeV$^2$/fm (upper panels) and $\hat q = 5$ GeV$^2$/fm (lower panels). The left panels correspond to the first splitting being $q\rightarrow qg$ and the right panels to $g\rightarrow gg$. The $a_{\rm med}/a$ ratio is independent of the second splitting kinematics. The first observation to make is that medium interactions suppress the strength of spin correlations in the bulk of phase-space. Namely, $a_{\rm med}/a$ remains $\leq 1$ and, when increasing $\hat q$, the phase space band where this modulation suppression is significant gets wider and, for most of phase space, stronger. In particular phase-space regions, it is even possible that the medium completely washes out the spin correlations effect.
Medium effects only vanish for very collinear splittings, $\theta_1\lesssim 10^{-2}$. The $(z_1,\theta_1)$-dependence of the suppression can be traced back to the formation time of the first splitting given by $\tau_1^{-1} \sim z_1(1-z_1)E_a\theta_1^2$. Then, for $\tau_1 > L$ or $\tau_1 \ll L$ the effect becomes negligible, as expected. Finally, when comparing both channels, we observe that the most significant difference occurs at $z_1 \gtrsim 0.5$. In turn, for $z_1 \ll 1$ no channel-dependence in $a_{\rm med}/a$ is observed since, in this limit, $P_{gq}(z_1) \approx P_{gg}(z_1) \approx 2/z_1$ and the same equality holds for the process-dependent functions in the medium modification in Eq.~\eqref{eq:process_dep_hf}.

\begin{figure}
     \centering
         \includegraphics[width=0.49\textwidth]{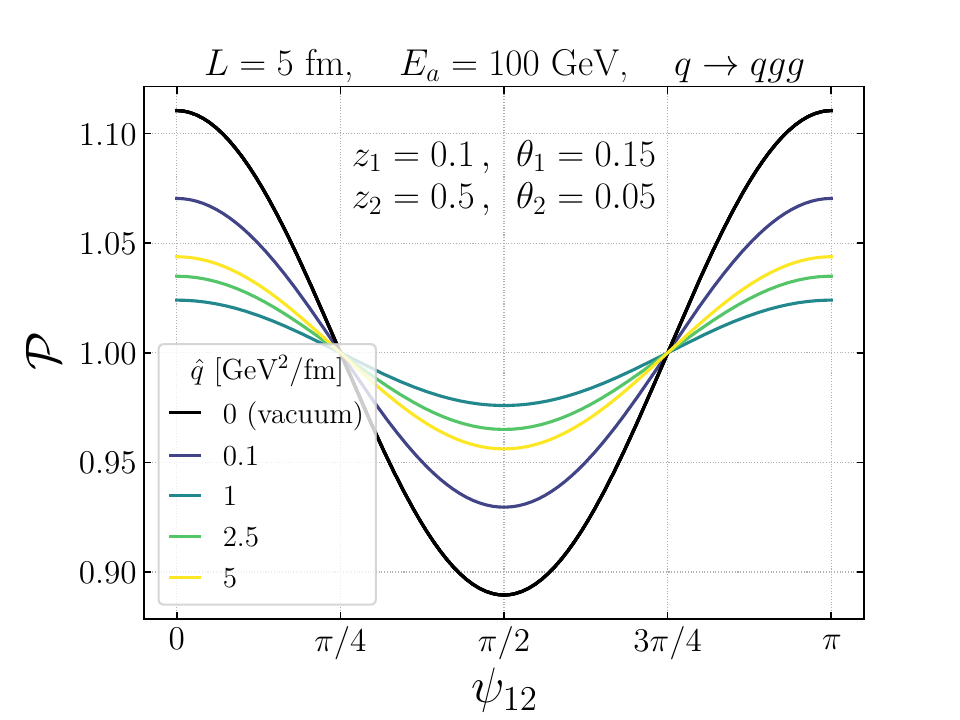}
         \includegraphics[width=0.49\textwidth]{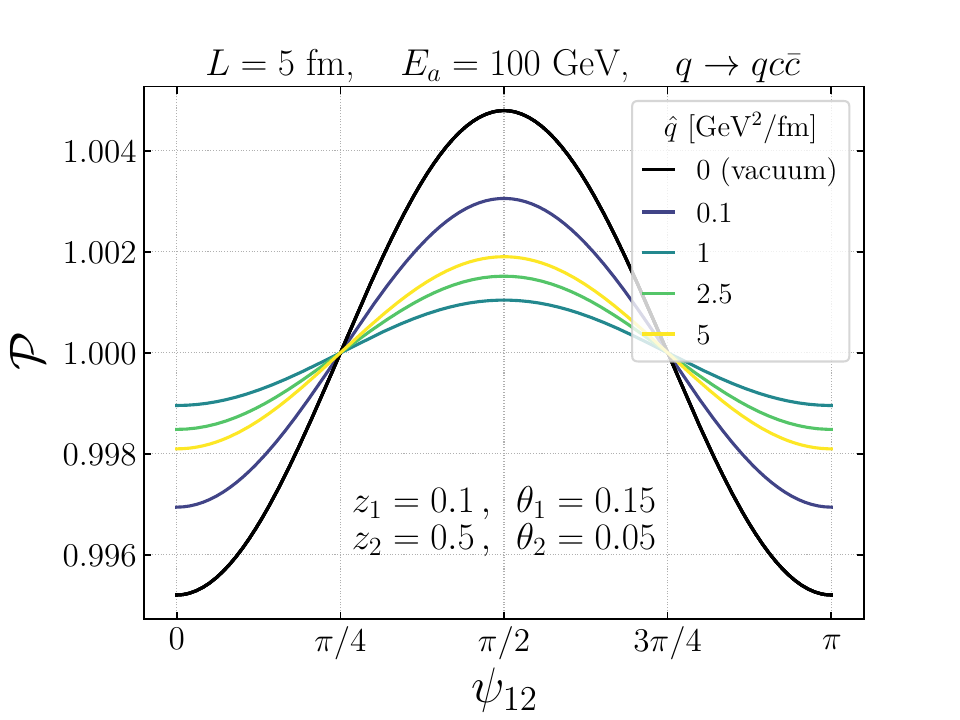}
         \\
         \includegraphics[width=0.49\textwidth]{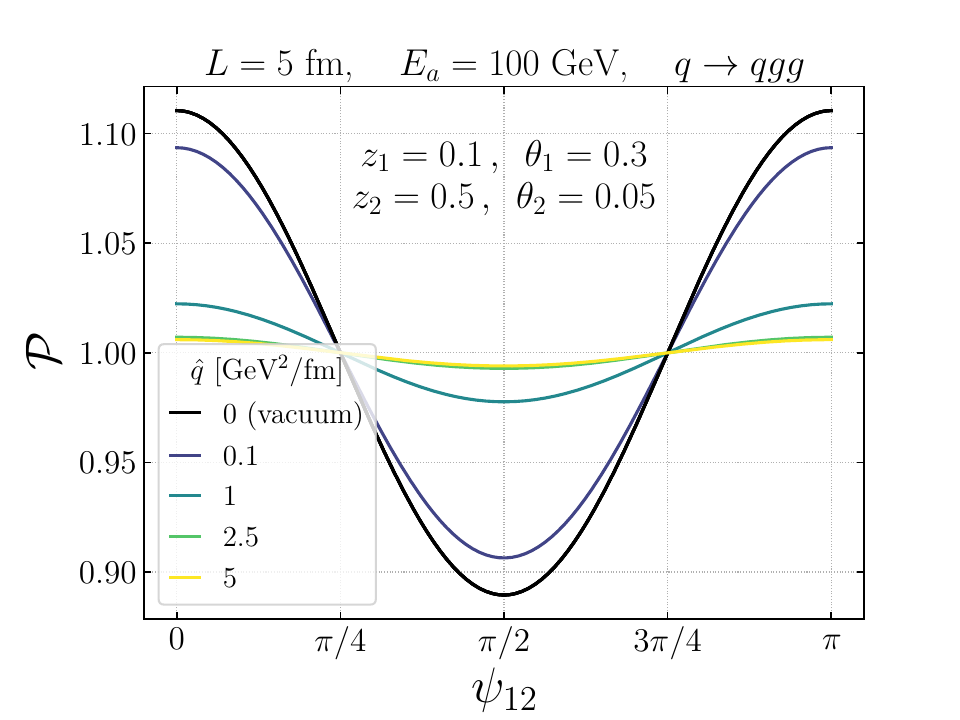}
         \includegraphics[width=0.49\textwidth]{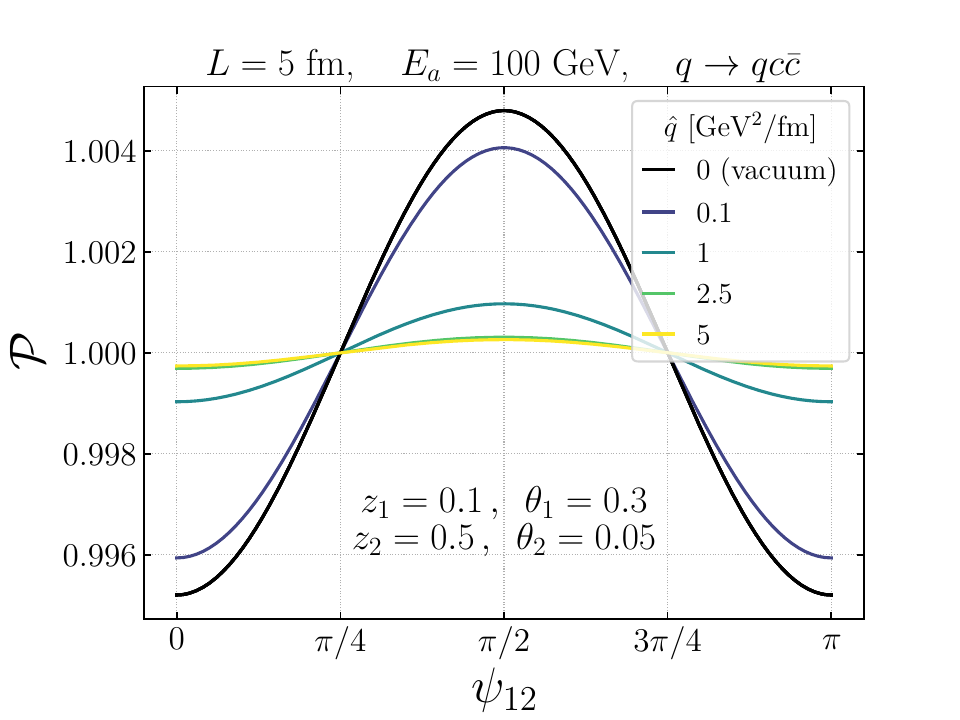}
     \caption{Evolution of the angular modulation in Eq.~\eqref{eq:p-def} with $\hat q$ as a function of $\psi_{12}$ for $\theta_1 = 0.15$ (top panels) and $\theta_1 = 0.3$ (bottom panels). The remaining kinematic variables are fixed to $z_1 = 0.1$, $z_2=0.5$ and $\theta_2 = 0.05$. The left panels correspond to $q\rightarrow qg \rightarrow qgg$ and the right panels to $q\rightarrow qg \rightarrow qc\bar c$. Note that for a massless second splitting (left panels) there is no dependence on $\theta_2$.}
     \label{fig:psi_isotropic}
\end{figure}

Next, we evaluate the in-medium cross section stripped off both the vacuum cross-section and the $(1+F_{\rm med})$ factor, i.e., we study
\begin{equation}\label{eq:p-def}
\mathcal{P}(z_1,\theta_1,z_2,\theta_2,\psi_{12})\equiv 1 + a_{\rm med} \cos(2\psi_{12}),
\end{equation}
as a function of $\psi_{12}$ in Fig.~\ref{fig:psi_isotropic}. We pick $(z_1,\theta_1)$ values for which medium modifications are strongest in Fig.~\ref{fig:A2_over_A1_iso}. The value of $z_2=0.5$ is also chosen so as to maximize the azimuthal modulations, see Fig.~\ref{fig:a_vac}. Finally, we fix $\theta_2=0.05<\theta_1$ to stay within the strongly-ordered approximation. Let us first discuss the vacuum results (black lines). As has been observed in the literature, $g\to gg$ and $g\to q\bar q$ splittings feature a correlation with opposite signs. Namely, for $q\to qgg$ splittings (left panels) $\mathcal{P}$ is largest when the two splittings are in the same plane ($\psi_{12}=0$), while for $g\to q\bar q$ (right panels) the spin correlations are largest when the splitting planes are perpendicular. In the massless case, such a modulation is much more pronounced in the $g \to q\bar q$ than for $g \to gg$ branchings, see for instance Refs.~\cite{Richardson:2018pvo, Karlberg:2021kwr}. However, when considering quarks to be massive, in particular setting $m_2=m_c$, we get an almost negligible modulation for the explored phase-space points. This is because for the chosen kinematics $\tilde\theta_{m,2}\approx 5$, i.e., the splitting occurs deep inside the dead cone. This weak modulation of $\mathcal{P}$ is consistent with the evolution of $a$ with $\tilde \theta_{m,2}$ as displayed in Fig.~\ref{fig:a_vac}. As for the medium results (coloured curves), we find the modulation to be suppressed in agreement with Fig.~\ref{fig:A2_over_A1_iso}. Further, the evolution of $\mathcal{P}$ with $\hat q$ depends on the phase-space point under consideration. For instance, when considering moderately large opening angles such as $\theta = 0.3$ (bottom panels), there is a monotonic trend - larger $\hat q$ implies a larger suppression of the modulation. For $\theta = 0.15$ (top panels), however, the trend is non-trivial - for $\hat q \sim 1$ GeV$^2$/fm we see a turning point for which the previous trend is reversed. More concretely, for $\hat q = 5$ GeV$^2$/fm the modulation is less suppressed than for $\hat q = 1$ GeV$^2$/fm. This can be traced back to Fig.~\ref{fig:A2_over_A1_iso}, where an increase in $\hat q$ is not only responsible for a widening of the band where suppression is significant, but also for a slight shift of the band's boundary. Therefore, we conclude that the angular modulation in the medium is always weaker than in vacuum and its magnitude depends both on the medium properties and the splitting kinematics. 


\subsection{Anisotropic medium}\label{subsec:anisotropic}
\begin{figure}
     \centering
         \includegraphics[width=\textwidth]{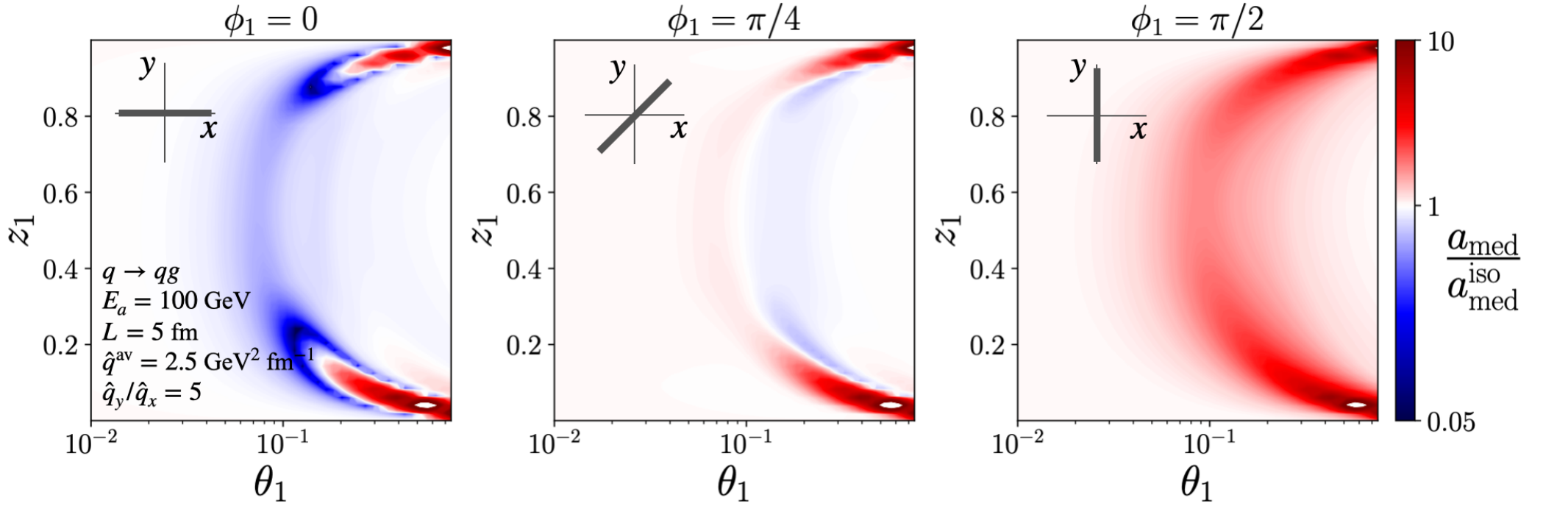}
     \caption{Modulation amplitude $a_{\rm med}$ for an anisotropic medium, divided by the isotropic result with $\hat q = \hat q^{\rm av.}$ (upper panels) as a function of the first splitting's kinematics $z_1$ and $\theta_1$, for the process $q\rightarrow qg$. The three columns correspond to three different azimuthal orientations of first splitting's plane: $\phi_1 = 0,\,\pi/4,\,\pi/2$.}
    \label{fig:amed_anisotropic}
\end{figure}
\begin{figure}
         \centering
         \includegraphics[width=\textwidth]{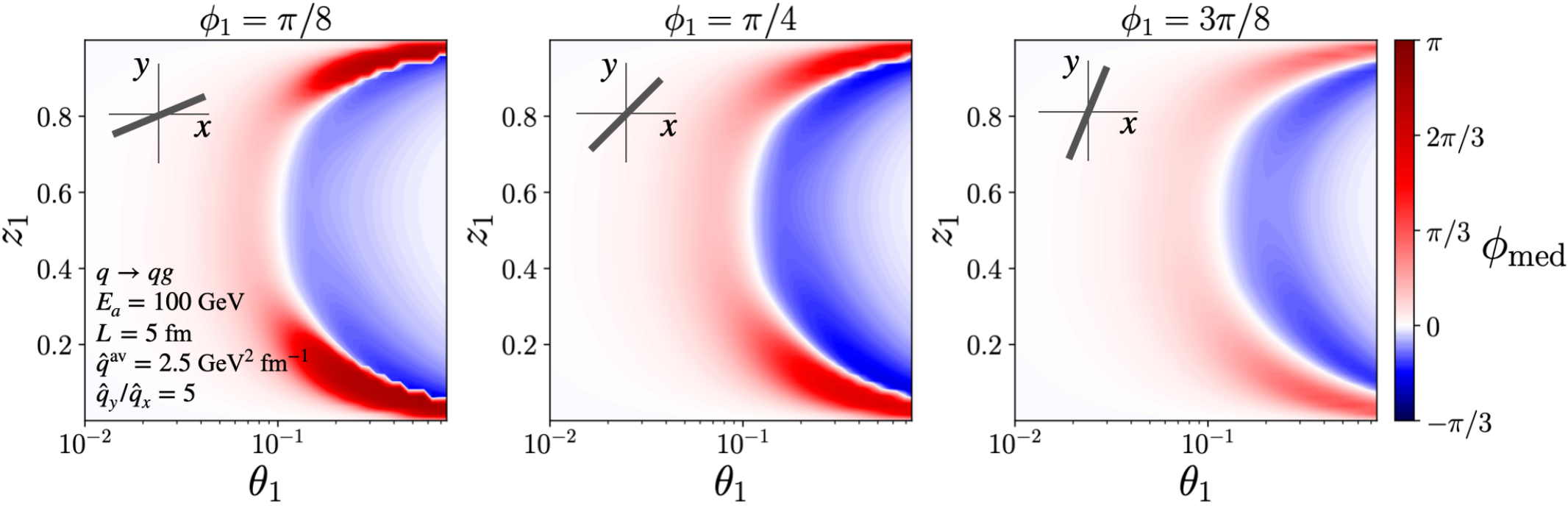}
         \caption{Same as Fig.~\ref{fig:amed_anisotropic} but for the phase shift $\phi_{\rm med}$. The values of $\phi_1$ differ with respect to Fig.~\ref{fig:amed_anisotropic} since $\phi_{\rm med}=0$ for $\phi_1=0,\,\pi/2$. Here we choose $\phi_1= \pi/8,\,\pi/4,\,3\pi/8$.}
    \label{fig:phimed_anisotropic}
\end{figure}
We now discuss the case of an anisotropic medium fixing $\hat q^{\rm av} = \frac{1}{2}(\hat q_x + \hat q_y) = 2.5$ GeV$^2$/fm and the anisotropy ratio to $\hat q_y/\hat q_x = 5$. The medium length is fixed to $L=5$ fm. Since the dependence on the first channel has been analyzed in the previous section, see Figs.~\ref{fig:A2_over_A1_iso},\ref{fig:psi_isotropic}, we focus on a single process: $q\rightarrow qg \to q gg$. The results are presented as a function of the first splitting's kinematics $(z_1,\theta_1,\phi_1)$ with fixed parent's energy $E_a=100$ GeV. 

In Fig.~\ref{fig:amed_anisotropic} we plot the ratio $a_{\rm med }/a_{\rm med}^{\rm iso}$ in the ($z_1,\theta_1$)-plane. This quantity measures the anisotropy-induced modifications on the amplitude of the spin correlations signal. First, we notice that $a_{\rm med }/a_{\rm med}^{\rm iso}$ deviates from unity in a band of phase space which roughly coincides with the one in Fig.~\ref{fig:A2_over_A1_iso}. As discussed, the shape of this band is controlled by the hierarchy between formation time $\tau_1$ and medium length $L$. Depending on the azimuthal angle $\phi_1$, $a_{\rm med}$ is enhanced (red) or suppressed (blue) in the bulk of phase space with respect to the isotropic case $a_{\rm med}^{\rm iso}$. In particular, for splitting planes orthogonal to the maximal anisotropy direction ($\phi_1 = 0$), one gets mostly a suppression, while for splitting planes aligned with this direction ($\phi_1 = \pi/2$) one gets an enhancement. Interestingly, for $\phi_1 = \pi/4$ one gets a very weak anisotropy-induced modification, which gets significant only for asymmetric, wide angle splittings.  

As discussed throughout the calculation presented in Sec.~\ref{sec:in_medium_spin_corr}, medium anisotropies lead to a unique signature on the spin correlations signal in terms of a phase-shift $\cos(2\psi_{12})\to \cos(2\psi_{12}+\phi_{\rm med})$, with  $\phi_{\rm med}$ defined in Eq.~\eqref{eq:phimed-def}. We show this phase-shift in Fig.~\ref{fig:phimed_anisotropic} as a function of ($z_1,\theta_1,\phi_1$). Since $\phi_{\rm med}$ vanishes exactly for $\phi_1=0,\,\pi/2$ (see Eq.~\eqref{eq:Ai_coeff_aniso}) we choose a different set of $\phi_1$ values with respect to Fig.~\ref{fig:amed_anisotropic}. We find that the phase-shift is more pronounced for splitting planes close-to-orthogonal to the maximal anisotropy direction $\phi_1 = \pi/8$. In fact, for specific regions of phase-space, e.g. $\theta_1> 0.1$ and $z_1<0.2$, we obtain $\phi_{\rm med} \approx \pi$ resulting in a sign-change in the azimuthal correlations. Finally, as one increasingly aligns the splitting plane with the maximal anisotropy ($\phi_1 = 3\pi/8$), $\phi_{\rm med}$ gradually takes smaller values.

\begin{figure}
     \centering
         \includegraphics[width=0.49\textwidth]{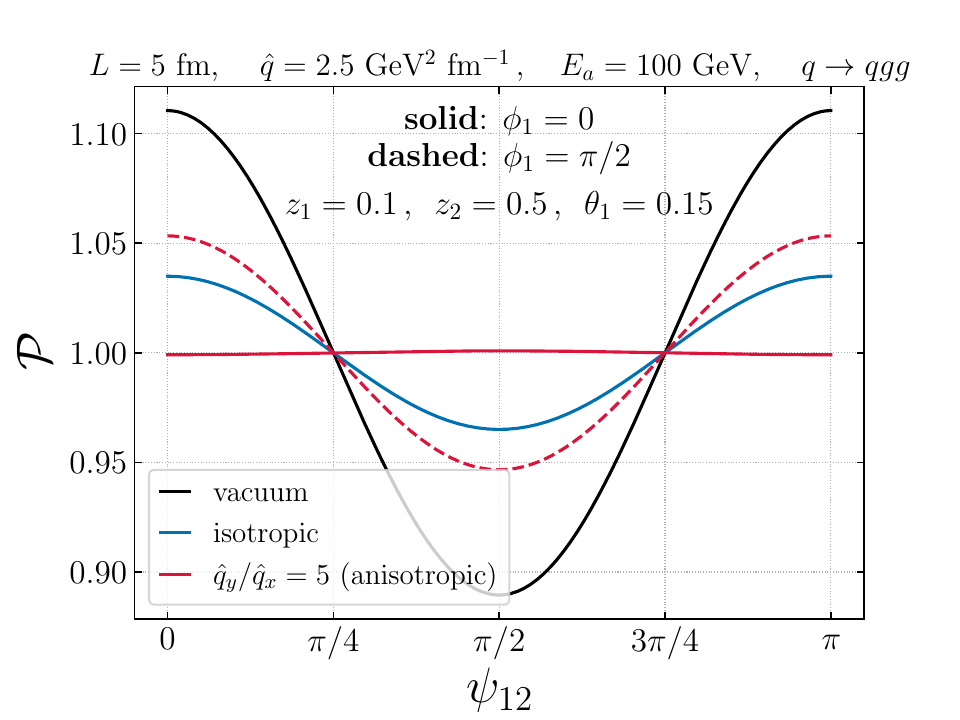}
         \includegraphics[width=0.49\textwidth]{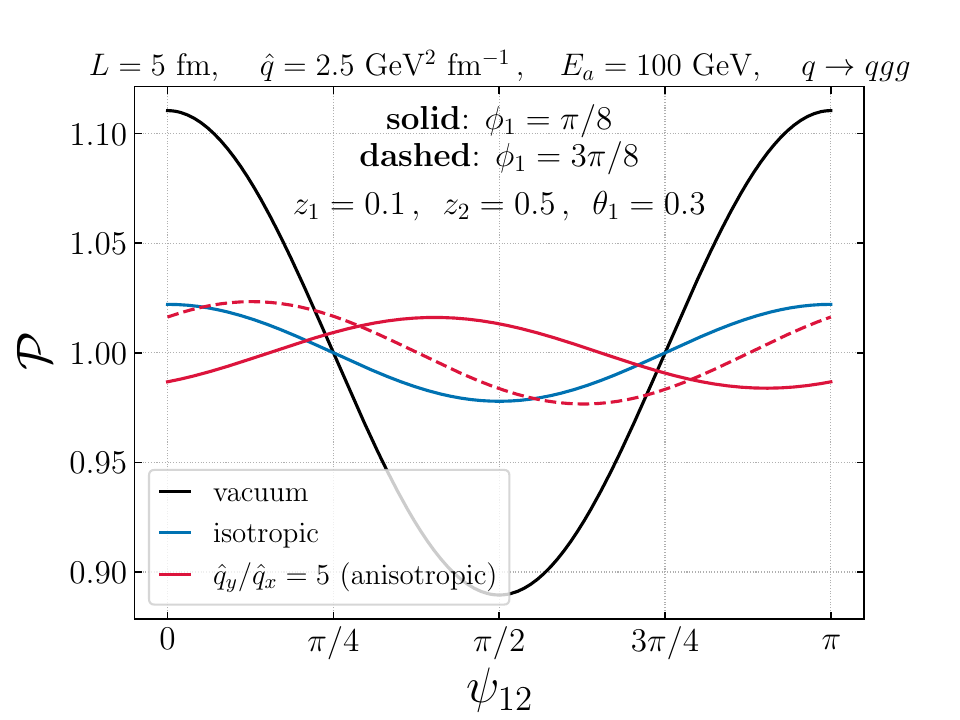}
     \caption{Angular modulation in Eq.~\eqref{eq:p-def} as a function of $\psi_{12}$ for vacuum (black), isotropic medium (blue) and anisotropic medium (red). The process is fixed to be $q\rightarrow qg \rightarrow qgg$, so there is no dependence on $\theta_2$. The energy fractions are fixed to $z_1 = 0.1$ and $z_2 = 0.5$, while $\theta_1 = 0.15$ on the left panel and $\theta_1 = 0.3$ on the right panel. The solid and dashed curves correspond to the first splitting's plane being more aligned with the $x$ and $y$ axis, respectively. In the left plot the medium anisotropy causes only a modulation change, while on the right one it induces only a phase-shift.}
    \label{fig:psi_anisotropic}
\end{figure}

To conclude, we now evaluate the in-medium modification to the cross-section, excluding the $(1+F_{\rm med})$ factor:
\begin{align}
\label{eq:p-anisotropic}
    \cP(z_1,\theta_1,\phi_1,z_2,\psi_{12}) \equiv 1 + a_{\rm med}\cos(\psi_{12}+\phi_{\rm med}).
\end{align}
Similarly to Fig.~\ref{fig:psi_isotropic}, we fix the energy fractions to $z_1=0.1$ and $z_2=0.5$ and choose $\theta_1=0.15$ (left panel) and $\theta_1=0.3$ (right panel). Since we only focus on the result for a massless second splitting $\cP$ does not depend on $\theta_2$. We plot Eq.~\eqref{eq:p-anisotropic} as a function of $\psi_{12}$ in Fig.~\ref{fig:psi_anisotropic} for vacuum (black curve), isotropic medium (blue curve) and anisotropic medium with $\hat q_y/\hat q_x = 5$ (red curves). The solid curve corresponds to the first splitting's plane close to (right panel, $\phi_1=\pi/8$) or exactly (left panel, $\phi_1 = 0$) orthogonal with maximum anisotropy direction, while the dashed curve corresponds to the first splitting's plane close to (right panel, $\phi_1 = 3\pi/8$) or exactly (left panel, $\phi_1 = \pi/2$) aligned with this direction. The left panel showcases the impact of medium anisotropies on the spin modulation signal purely at the magnitude ($a_{\rm med}$) level since for $\phi_1 = 0,\,\pi/2$ one has $\phi_{\rm med}=0$. For $\phi_1=0$ (solid) we observe $\mathcal{P}=1$ or, equivalently, $a_{\rm med}=1$ in Eq.~\eqref{eq:p-anisotropic}, while for $\phi_1 = \pi/2$ (dashed) there is a modulation enhancement with respect to the isotropic case. Thus, for certain regions of phase space, anisotropy can fully wash out azimuthal correlations even if they're present for the same region in the case of an isotropic medium.
The right panel highlights instead the phase-shift in $\cP$ since for this choice of ($z_1,\theta_1,\phi_1$) the value of $a_{\rm med}/a^{\rm iso}_{\rm med}$ remains close to $1$ as can be seen by comparing the blue line with the red, dashed line in the right panel of Fig.~\ref{fig:psi_anisotropic}. For $\phi_1=\pi/8$ we observe that the anisotropic result for $\cP$ is shifted almost 90 degrees with respect to the vacuum and isotropic results ($\phi_{\rm med} \approx \pi$), in agreement with the findings presented in Fig.~\ref{fig:phimed_anisotropic}. We discuss some of the phenomenological challenges related to measuring this phase-shift in the next section.

\section{Conclusions}
\label{sec:conclusions}

Quantum mechanical interferences driven by colour and spin degrees of freedom are known to break the independent emission picture of QCD radiation. Modifications to the colour interference pattern in the context of jets in heavy-ion collisions have been studied in a series of papers~\cite{Mehtar-Tani:2010ebp,Mehtar-Tani:2011hma,Casalderrey-Solana:2011ule,Mehtar-Tani:2011vlz,Mehtar-Tani:2011lic,Armesto:2011ir,Mehtar-Tani:2012mfa,Casalderrey-Solana:2012evi,Calvo:2014cba,Barata:2021byj,Abreu:2024wka,Andres:2025prc,Kuzmin:2025fyu}. 
In this work, we have initiated the study of collinear spin correlations in QCD jets that propagate through a quark-gluon plasma. To that end, we have computed the fully differential cross-section to produce a nearly on-shell gluon inside the QGP ($a\to b\,g$) and its subsequent branching outside of the medium ($g\to c\,d$).
The calculation includes mass effects and is performed in the strongly ordered \textit{quasi}-collinear limit. The main result of this paper is the medium-modified cross section for the process $a\to b\,g\to b\,c\,d$ given in Eq.~\eqref{eq:xsec_med_full}. Besides the well-known modification of the total rate of emissions encoded in the $F_{\rm med}$ factor, we observe some new striking effects. 
In vacuum, the polarisation of the intermediate gluon induces an azimuthal correlation between the production and the decay planes. In a dense medium, under our approximations, we find that both the amplitude of this modulation ($a_{\rm med}$) and its phase ($\phi_{\rm med}$) change.
These modifications have a non-trivial parametric dependence on the kinematics and partonic flavour of the first splitting as well as on the medium properties. For instance, $a_{\rm med}$ is always smaller than the vacuum baseline but the magnitude of this depletion depends on the channel under consideration, the phase-space point as well as on the medium density, see Figs.~\ref{fig:A2_over_A1_iso},\ref{fig:amed_anisotropic}. For specific regions of phase space, the medium can completely suppress the azimuthal modulation. The phase-shift, $\phi_{\rm med}$, only appears when considering anisotropic media since they introduce a preferred direction to which the spin correlation couples. Remarkably, we have shown that for certain phase-space points the medium leads to a 90 degree shift with respect to the vacuum modulation.

There are several caveats when trying to translate the findings of this exploratory, theory study to a realistic measurement at the LHC. Some of these caveats are already present in the proton-proton case. In particular, the observable we have studied, namely the $\psi_{12}$-distribution, was first proposed in Refs.~\cite{Richardson:2018pvo,Karlberg:2021kwr}. There, it was shown that the size of the modulation in a full parton shower decreases with respect to the fixed-order calculation at $\mathcal{O}(\alpha^2_s)$. This is due to two effects: (i) the different sign in the modulation featured by $g\to gg$ and $g\to q\bar q$ splittings and (ii) the presence of other branchings not mediated by gluons. These two effects result into the overall spin effect mostly canceling out when considering inclusive jets, as observed in preliminary data from the CMS collaboration~\cite{CMS:2025awh}. In the medium case, the phase difference between $g\to gg$ and $g\to q\bar q$ splittings in the $\psi_{12}$-distribution competes with the anisotropy-induced phase-shift $\phi_{\rm med}$ that we have studied in Figs.~\ref{fig:phimed_anisotropic} and \ref{fig:psi_anisotropic}. A possible way out, that has been pursued in the CMS analysis, is to apply heavy-flavour tagging techniques to isolate $g\to q\bar q$ splittings and thus obtain a cleaner spin-correlation signal. Such measurement in the heavy-ion case is apriori statistics limited but it remains to be studied whether it could become feasible in the high-luminosity phase of the LHC~\cite{Citron:2018lsq}. Other proposals to experimentally access spin-dependent quenching effects or medium anisotropies have been presented in Refs.~\cite{Barata:2023zqg,Barata:2024bqp,Barata:2025uxp,Yao:2025ldx}.


From a theoretical perspective, there are at least three directions that we would like to pursue to extend the present calculation. The most pressing one, although not straightforward, is to explore the case in which the second splitting takes place inside the medium. A complete calculation of the $1\to 3$ matrix-element including multiple scatterings with the medium is really challenging. As an intermediate step, one could think of using the $1\to 3$ results in the collinear limit presented in Ref.~\cite{Fickinger:2013xwa} where only one scattering between the propagating parton and the medium constituents was considered. 
Another aspect is how to iterate the calculation that we have presented for multiple branchings in the context of the cascade approximation~\cite{Fickinger:2013xwa, Barata:2025fzd}. In other words, it would be very interesting to formulate the medium analog of the Collins-Knowles algorithm~\cite{Collins:1987cp,Knowles:1987cu}. Finally, we look forward to studying the imprint of medium-induced spin flips on azimuthal correlations between splittings by including helicity-dependent terms in the medium averages as was done in Ref.~\cite{Cougoulic:2020tbc}. 

The calculation presented in this paper has revealed new qualitative effects in the azimuthal structure of jets in heavy-ion collisions. These results represent a clear step forward in our understanding of in-medium jet physics and we look forward to investigate further into their phenomenological implications in future works.

\acknowledgments
We thank Melissa van Beekveld for insightful discussions on the role of mass effects in spin correlations and João Barata for comments on the manuscript. J.M.S. has been supported by MCIN/AEI (10.13039/501100011033) and ERDF (grant PID2022-139466NB-C21) and by Consejería de Universidad, Investigación e Innovación, Gobierno de España and Unión Europea – NextGenerationEU under grant AST22 6.5. ASO is supported by the Ramón y Cajal program under grant RYC2022-037846-I and from ERDF (grant PID2024-161668NB-100).

\appendix
\section{Calculation details}
\label{app:details}
\subsection{Kinematics and phase-space parametrization}\label{app:kinematics_phase_space}
In the main text we consider any $1\to 3$ process such that $a\to b\,g \to b\,c\,d$. We first introduce the light-cone energy fractions $z_i$ and the relative transverse momentum of each, $1\to 2$ splitting 
\begin{subequations}\label{eq:relative_momenta_def}
\begin{align}
    & z_1 = p_g^+ / p_a^+\,, \quad \q_1 = (1-z_1)\p_g - z_1\p_b \\
	& z_2 = p_c^+ / p_g^+\,, \quad \q_2 = (1-z_2)\p_c - z_2\p_d
\end{align}
\end{subequations}
where we denote $\p_g = \p_c+\p_d$ and the total transverse momentum is  $\p_a = \p_b+\p_c+\p_d$. Physically, $z_1$ is the energy fraction of the intermediate gluon, and $z_2 $ is the energy fraction of the final-state particle with momentum $p_c$. It's also convenient to define
\begin{subequations} \label{eq:omega-def}
\begin{align}
\omega_a = z_1(1-z_1)p_a^+ \,,\\
\omega_g = z_2(1-z_2)p_g^+ \,.
\end{align}
\end{subequations}
We wish to make the change of coordinates $\{\p_a,\q_{1},\q_{2}\} \rightarrow \{\p_a,\theta_{1},\theta_{2},\phi_{1},\psi_{12}\}$, the definitions of which will become clear in the following text. It is useful to first invert the relations for the transverse momenta in Eq.~\eqref{eq:relative_momenta_def}
\begin{subequations}\label{eq:inverse_relative_momenta_def}
\begin{align}
	& \p_b = (1-z_1)\p_a-\q_{1}\,,\\
	& \p_c = z_1(1-z_2)\p_a+ (1-z_2) \q_{1} -\q_{2}\,, \\
	& \p_d = z_1z_2\p_a + z_2\q_{1}+ \q_{2}\,.
\end{align}
\end{subequations}
and to write the $z$-component of each momentum in terms of energy and transverse momentum i.e. $p_{iz} = \sqrt{E_i^2- \p_i^2 - m_i^2}$, such that each cartesian three vector is defined in terms of $\p_a$, $\q_{1}$, $\q_{2}$ and $m_i$. In what follows we work in the quasi-collinear limit,  that is we scale all transverse momenta as $\p\rightarrow \lambda \p$ and masses as $m_i \rightarrow \lambda m_i$ and keep the first non-vanishing term in a power series of $\lambda\ll1$. 

Then, defining the opening angles of each splitting through the dot product of appropriate three-vectors one finds
\begin{subequations}\label{eq:opening_angles_app}
\begin{align}
	& \theta_{1}^2/2 = 1 - \frac{\vec p_b\cdot\vec p_g}{|\vec p_b||\vec p_g|} =  \q_{1}^2/\omega_a^2 + \mathcal{O}(\lambda^4) \,,\\
	& \theta_{2}^2/2 = 1 - \frac{\vec p_c\cdot\vec p_d}{|\vec p_c||\vec p_d|} = \q_{2}^2/\omega_g^2  + \mathcal{O}(\lambda^4)\,.
\end{align}
\end{subequations}
%
These expressions for the splitting opening angles are valid up to terms of order $\mathcal{O}(\lambda^4)$, where $\lambda^4$ represents any product of four transverse momenta $\p$ and/or masses $m_i$.
To relate the azimuthal angles $\phi_{1}$ and $\phi_{2}$ to the angle between the planes spanned by the splittings, which we denote by $\psi_{12}$, we start with the definition of this quantity as the angle between the normal vectors to each plane. The normal three-vectors of unit norm are defined as the following cross product
\begin{align}
	& \vec n_{1} = \frac{\vec p_a \times \vec p_g}{|\vec p_a \times \vec p_g|}\,,\quad \vec n_{2} = \frac{\vec p_c \times \vec p_d}{|\vec p_c \times \vec p_d|}\,,
\end{align}
so that $\cos(\psi_{12})$ is defined as 
\begin{equation}
    \cos(\psi_{12}) = \vec n_{1}\cdot\vec n_{2}\,,
\end{equation}
%
which can now be written as a function of $\p_a$, $\q_{1}$, $\q_{2}$ and $m_i$ exactly as was done for the opening angles, using Eq.~\eqref{eq:inverse_relative_momenta_def}. By keeping only terms up to $\mathcal{O}\left(\lambda^2\right)$, where $\lambda^2$ is any quadratic product of $\p$ and $m_i$, one arrives at 
\begin{equation}\label{eq:phi_to_psi}
	\cos\psi_{12} = \frac{\q_{1}\cdot\q_{2}}{|\q_1||\q_2|} + \mathcal{O}(\lambda^2) \implies \psi_{12} = \phi_{2} - \phi_{1}\,.
\end{equation}

By using the relations in Eq.~\eqref{eq:opening_angles_app} together with Eq.~\eqref{eq:phi_to_psi}, we can now write the Jacobian for the change of coordinates $\{\p_a,\q_{1},\q_{2}\} \rightarrow \{\p_a,\theta_{1},\theta_{2},\phi_{1},\psi_{12}\}$
\begin{align}
\label{eq:jacobian}
	d^2\p_a d^2\q_{1}d^2\q_{2} & = \omega_a^2\omega_g^2\theta_{1}\theta_{2}/4\left(d^2\p_a d\theta_{1}d\theta_{2}d\phi_{1}d\psi_{12}\right)\nn
	& = d^2\p_a d\theta_{1}d\theta_{2}d\phi_{1}d\psi_{12} \frac{z_1^4(1-z_1)^2z_2^2(1-z_2)^2(p_a^+)^4\theta_{1}\theta_{2}}{4}.
\end{align}
This Jacobian becomes relevant when promoting splitting amplitudes $\mathcal{M}$ to differential cross-sections, by including the appropriate phase-space element, i.e.
\begin{align}
\label{eq:phase-space-element}
	2p_b^+2p_c^+2p_d^+(2\pi)^9\frac{d\sigma}{dp_b^+dp_c^+dp_d^+d^2\p_bd^2\p_cd^2\p_d} = \frac{1}{2}\sum_{\rho_i,\lambda_i} \cM\cM^{\dagger}\,.
\end{align}
where the sum runs over colours ($\rho_i$) and spin states ($\lambda_i$). The factor $1/2$ accounts for the average over the initial parton spin state that we have denoted $\lambda_a$ in the main text. Let us rewrite the phase-space element in Eq.~\eqref{eq:phase-space-element} in terms of the relative transverse momenta $\q_i$ and energy fractions $z_i$
\begin{align}
\label{eq:phase-space-two}
2z_1(1-z_1)2z_2(1-z_2)(2\pi)^6\frac{d\sigma}{d^3\Omega_a\,dz_1\,d z_2\,d^2\q_{1}d^2 \q_{2}} = \frac{1}{2}\sum_{\rho_i,\lambda_i} \cM\cM^{\dagger}\, ,
\end{align}
where
\begin{align}
d^3\Omega_a = \frac{dp_a^+}{(2\pi)2p_a^+}\frac{d^2\p_a}{(2\pi)^2}\, ,
\end{align}
corresponds to the phase-space element of the total momentum of the system.
Then, by using Eqs.~\eqref{eq:jacobian} and \eqref{eq:phase-space-two}, we can re-write Eq.~\eqref{eq:phase-space-element} as 
\begin{align}
\label{eq:phase-space-final}
	\frac{16(2\pi)^6}{z_1^3(1-z_1)z_2(1-z_2)\theta_{1}\theta_{2}}\frac{d\sigma}{dz_1\,d z_2\,d\theta_{1}d\theta_{2}d\phi_{1}d\psi_{12}} =  \frac{1}{2}\sum_{\rho_i,\lambda_i}\int\, d^3\Omega_a (p_a^+)^4 \cM\cM^{\dagger} \,.
\end{align}
To keep the notation compact, we define
\begin{equation}
\label{eq:phips-def}
d^6\Omega_{\rm PS} \equiv dz_1\,d z_2\,d\theta_{1}d\theta_{2}d\phi_{1}d\psi_{12}\,,
\end{equation}
and express the fully differential cross-section as $d\sigma/d^6\Omega_{\rm PS}$.
\subsection{Spin-dependent QCD vertices}\label{app:vertex}
Consider the QCD splitting $i (p_0, \lambda_0) \rightarrow j (p_1, \lambda_1) k (p_2, \lambda_2)$, where $p_i$ are the particle momenta and $\lambda_i = \pm 1$ are their spin states. The spin-dependent vertices can be directly calculated from the following expressions
\begin{subequations}
\begin{align}
	&V_{q\rightarrow gq}^{\lambda_0 \lambda_1 \lambda_2}(p_1,p_2) = \bar u^{\lambda_2}(p_2) \slashed{\epsilon}^{\ast, \lambda_1}(p_1)u^{\lambda_0}(p_0) \\
	&V_{g\rightarrow q\bar q}^{\lambda_0 \lambda_1 \lambda_2}(p_1,p_2) = \bar u^{\lambda_1}(p_1) \slashed{\epsilon}^{\lambda_0}(p_0)v^{\lambda_2}(p_2)\,,
    \\
	 &V_{g\rightarrow gg}^{\lambda_0 \lambda_1 \lambda_2}(p_1,p_2) = \epsilon^{\lambda_0}_{\alpha}(p_0)\epsilon^{\ast, \lambda_1}_{\beta}(p_1)\epsilon^{\ast,\lambda_2}_{\gamma}(p_2)[g^{\alpha\beta}(p_0+p_1)^{\gamma} - g^{\beta\gamma}(p_1-p_2)^{\alpha}-g^{\gamma\alpha}(p_0+p_2)^{\beta}]\,,
    \end{align}
\end{subequations}
where $p_0 = p_1+p_2$. The states $\lambda_i$ are defined as twice the spin projection along $z$ for quarks and as the circular polarisation state for gluons. The gluon polarisation vector is defined in light-cone gauge as
\begin{align}\label{eq:pol_vector}
	\epsilon^{+,\lambda}(p) = 0\,,\quad  \epsilon^{-,\lambda}(p) = \frac{\bepsilon^{\lambda}\cdot\p}{p^+}\,,\quad \bepsilon^{\lambda} = \frac{1}{\sqrt{2}}(1,i\lambda)\,.
\end{align}
By using this polarisation vector together with the canonical Dirac spinors, these vertices read (see e.g.~\cite{Pauli:2000gw, Kovchegov:2012mbw,Lappi:2016oup, Richardson:2018pvo})
\begin{subequations}\label{eq:vertex_all_channels}
\begin{align}
	  &V_{q\rightarrow gq}^{\lambda_0 \lambda_1 \lambda_2}(z,\q) =\frac{2 \delta^{\lambda_0 \lambda_2}}{z\sqrt{1-z}}(\delta^{\lambda_0, \lambda_1} + (1-z)\delta^{\lambda_0, -\lambda_1})(\q\cdot \bepsilon^{\ast,\lambda_1}) - \lambda_0\delta^{\lambda_0,-\lambda_2}\delta^{\lambda_0,\lambda_1}\frac{\sqrt{2} z m}{\sqrt{1-z}}\, , \\
	  &V_{g\rightarrow q\bar q}^{\lambda_0 \lambda_1 \lambda_2}(z,\q) =\frac{2\lambda_1 \delta^{\lambda_1,\lambda_2} }{\sqrt{z(1-z)}}(z\delta^{\lambda_0,\lambda_1} - (1-z)\delta^{\lambda_0, -\lambda_1})(\q\cdot \bepsilon^{\lambda_0}) -\delta^{\lambda_1, -\lambda_2}\delta^{\lambda_0, \lambda_1} \frac{\sqrt{2}m}{\sqrt{z(1-z)}}\, , \\
	&V_{g\rightarrow gg}^{\lambda_0 \lambda_1 \lambda_2}(z,\q) = 2\left(\frac{\delta^{\lambda_0, \lambda_1}\delta^{\sigma,\lambda_2}}{1-z} - \delta^{\lambda_1, -\lambda_2}\delta^{\sigma, -\lambda_0} + \frac{\delta^{\lambda_0,\lambda_2}\delta^{\sigma, \lambda_1}}{z}\right)(\q\cdot \bepsilon^{\ast,\sigma})\,,
     \end{align}
\end{subequations}
where $z = p_1^+/p_0^+$ corresponds to the energy fraction of particle $j$ in the splitting and $\q = (1-z)\p_1 - z\p_2$.  Note that the dot product between the relative transverse momentum and the gluon polarisation vector carries information about the azimuthal angle, since one can write it as
\begin{equation}
	\q\cdot\bepsilon^{\lambda} = \frac{|\q|}{\sqrt{2}}e^{i \lambda\phi}\,.
\end{equation}
This allows to factor out the azimuthal dependence in each vertex as a single overall phase~\cite{Richardson:2018pvo, Karlberg:2021kwr}
\begin{align}\label{eq:vertex-and-phi}
	V^{\lambda_0 \lambda_1 \lambda_2 }(z,\q) = \tilde V^{\lambda_0 \lambda_1 \lambda_2}(z,|\q|)\, e^{i(\bar \lambda_0 - \bar \lambda_1 - \bar \lambda_2)\phi}\,,
\end{align}
where $\bar\lambda_i = \lambda_i$ if the particle is a gluon and $\bar \lambda_i = \pm \lambda_i / 2$ if the particle is a quark or anti-quark, respectively. In $\tilde V$ one replaces $\q\cdot\bepsilon\rightarrow |\q|/\sqrt{2}$ in $V$. This can be readily checked by plugging this phase in Eq.~\eqref{eq:vertex_all_channels}. 

In the main text, we wish to compute the cross-section for the process $a\rightarrow b\,g \rightarrow b\,c\,d$. At the scattering amplitude level, this involves a sum over the intermediate gluon's polarisations ($\lambda_g$) of a product of vertices,
\begin{align}\label{eq:vertex_term}
	\cA(\q_1,\q_2) \equiv \sum_{\lambda_g}V^{\lambda_g\lambda_c\lambda_d}_{g\rightarrow c d}(z_2,\q_2)V^{\lambda_a\lambda_g\lambda_b}_{a\rightarrow b g}(z_1, \q_1)\,,
\end{align}
while for the squared-matrix element we also need to sum over final state spin states $\lambda_b,\lambda_c,\lambda_d$ and average over the initial state one $\lambda_a$.
One can verify that, in vacuum, the product of any QCD vertex in Eq.~\eqref{eq:vertex_all_channels} and its complex conjugate is polarisation independent when considering a different polarisation for the off-shell gluon in the amplitude and complex conjugate amplitude, as was stated in Eq.~\eqref{eq:real_imag_F_M2}. For instance, for a $g\to q\bar q$ splitting with energy fraction $z$ and transverse momentum $\q$ we find
\begin{equation}
	\sum_{\lambda_q \lambda_{\bar q}} \tilde V^{\lambda_g \lambda_q \lambda_{\bar q}}_{g\rightarrow q\bar q}(z, \q)\tilde V^{\ast,\lambda'_g \lambda_q \lambda_{\bar q}}_{g\rightarrow q\bar q}(z, \q) = \frac{2}{z(1-z)}\left(\delta^{\lambda_g, \lambda'_g} (m_q^2 + \q^2P_{qg}(z))- \delta^{\lambda_g, -\lambda'_g}\, 2 z(1-z) \q^2 \right)\,,
\end{equation}
i.e., the functions attached to both $\delta$ tensors are independent of $\lambda_g$ and $\lambda'_g$. This implies, in particular, that there is no $\sin(2\psi_{12})$ modulation due to spin correlations (see Eq.~\eqref{eq:A_out_out}).
However, if we assume the first splitting $a\rightarrow b\,g$ to happen inside a QCD medium while the gluon splits in vacuum (see Eq.~\eqref{eq:real_imag_F_M2_medium}), one can show that polarisation dependent terms arise when there is some type of anisotropy in the transverse plane which breaks rotational invariance of the medium. To see this, consider the contribution coming from the term with $\lambda_g \neq \lambda'_g$ in the square of the $a\rightarrow b\,g$ vertex. Due to interactions with the medium, one writes this with different relative transverse momentum in the amplitude ($\bkappa$) and in the conjugate amplitude ($\bar\bkappa$),
%
\begin{align}\label{eq:squared_vertex_polarisation}
	\delta^{\lambda_g, -\lambda'_g}\sum_{\lambda_a\lambda_b}V^{\lambda_a\lambda_g \lambda_b}_{a\to gb}(z,\bkappa)V_{a\to gb}^{\ast, \lambda_a \lambda_g' \lambda_b}(z,\bar\bkappa)e^{i(\lambda_g-\lambda_g')\phi_{2}} \propto \delta^{\lambda_g, -\lambda'_g}e^{-i\lambda_g (\phi_\bkappa + \phi_{\bar\bkappa} - 2\phi_{2})}\,,
\end{align}
where $\bkappa = |\bkappa|(\cos\phi_\bkappa,\sin\phi_\bkappa)$ and the same for $\bar\bkappa$, and the angle $\phi_{2}$ is the azimuthal orientation of the final-state plane formed by $b$ and $g$. If the medium is rotationally invariant the dependence on $\phi_2$ in Eq.~\eqref{eq:squared_vertex_polarisation} vanishes. Also, the remaining function of $\bkappa$ and $\bar\bkappa$ against which Eq.~\eqref{eq:squared_vertex_polarisation} is integrated depends only on $\bkappa^2$, $\bar\bkappa^2$ and $\phi_\bkappa-\phi_{\bar\bkappa}$.\footnote{Note that this reasoning is only valid because we assumed the gluon splits in vacuum. Were this not true, then a dependence on $\psi_{12} = \phi_2-\phi_1$ would exist for the medium modification. Thus, the integrand could depend on, e.g., $(\phi_\bkappa-\psi_{12})+(\phi_{\bar\bkappa}-\psi_{12})$.} Hence, the phase $e^{i\lambda_g(\phi_\bkappa+\phi_{\bar\bkappa})}$ can be integrated out to yield a $\lambda_g$ independent result.

Finally, for all vertices in Eq.~\eqref{eq:vertex_all_channels}, one can show that the square of Eq.~\eqref{eq:vertex_term} entering the product of the amplitude and conjugate amplitude reads
\begin{align}\label{eq:AAdagger_all_channels}
    &  \frac{1}{2}\sum_{\lambda_a}\sum_{\lambda_b\lambda_c\lambda_d} \cA(\bkappa,\q_{2})\cA^{\dagger}(\bar\bkappa,\q_{2}) = \frac{4(m_2^2 + \q_{2}^2 P_{g\rightarrow cd}(z_2))}{z_2(1-z_2)z_1(1-z_1)}
    \times \Big[z_1^3 m_1^2 + P_{a\rightarrow gb}(z_1)\left(\bkappa\cdot \bar\bkappa\right)\nn
    &\mp\frac{4z_2(1-z_2)(1-z_1)}{z_1(m_2^2 + \p_{23}^2 P_{g\rightarrow cd}(z_2))}\left(\left(\q_{2}\cdot \bkappa\right)\left(\q_{2}\cdot \bar\bkappa\right)-\left(\q_{2}\times \bkappa\right)_z\left(\q_{2}\times \bar\bkappa\right)_z\right)\Big]
\end{align}
where $(\q_2\times \bkappa)_z = q_2^x\kappa^y-q_2^y\kappa^x$, $\bkappa$ and $\bar\bkappa$ correspond to integration variables in the medium calculation, and all other variables are defined in Eq.~\eqref{eq:relative_momenta_def}. The channel-dependence of Eq.~\eqref{eq:AAdagger_all_channels} is encoded in the factor $\pm$ which takes into account if the intermediate gluon splits into a $q \bar q$ pair ($+$) or a pair of gluons ($-$), and in $P_{ji}(z)$ which is the massless splitting function for the process $i\rightarrow j k$ stripped of its corresponding colour factor
\begin{subequations}
\label{eq:vac_splitting_functions}
\begin{align}
    & P_{q\rightarrow g q}(z) = \frac{1 + (1-z)^2}{z}\,, \\
	& P_{g\rightarrow q\bar q}(z) = z^2 + (1-z)^2\, ,\\
	& P_{g\rightarrow gg}(z) = 2\left(\frac{z}{1-z}+\frac{1-z}{z}+z(1-z)\right)\,.
\end{align}
\end{subequations}
Note that $m_1=m_a=m_b$ and and $m_2=m_c=m_d$. 

In the vacuum case Eq.~\eqref{eq:AAdagger_all_channels} can be further simplified by setting $\bkappa=\bar\bkappa=\q_1$ such that the squared amplitude reads
\begin{align}
      \frac{1}{2}\sum_{\lambda_a}\sum_{\lambda_b\lambda_c\lambda_d} \cA(\q_{1},\q_{2})\cA^{\dagger}(\q_{1},\q_{2}) &\underset{\text{vacuum}}{=}  \frac{4(m_2^2 + \q_{2}^2 P_{g\rightarrow cd}(z_2))}{z_2(1-z_2)z_1(1-z_1)}\times \Big[z_1^3 m_1^2 + P_{a\rightarrow gb}(z_1)\q_{1}^2\nn
    & \mp\frac{4z_2(1-z_2)(1-z_1)}{z_1(m_2^2 + \q_{2}^2 P_{g\rightarrow cd}(z_2))}\left(\left(\q_{2}\cdot \q_{1}\right)^2 -\left|\q_{2}\times \q_{1}\right|^2\right)\Big]\, .
\end{align}
By using Eq.~\eqref{eq:phi_to_psi}, we can re-write the scalar and vector products in terms of the azimuthal angle between the planes spanned by the two splittings:
\begin{align}\label{eq:A_out_out}
    & \frac{1}{2}\sum_{\lambda_a}\sum_{\lambda_b\lambda_c\lambda_d} \cA(\q_{1},\q_{2})\cA^{\dagger}(\q_{1},\q_{2}) \underset{\text{vacuum}}{=} \frac{4(m_2^2 + \q_{2}^2 P_{3g}(z_2))}{z_2(1-z_2)z_1(1-z_1)}\nn
    & \times \left[z_1^3 m_1^2 + P_{a\rightarrow gb}(z_1)\q_{1}^2\mp\frac{4z_2(1-z_2)(1-z_1)\q_{1}^2\q_{2}^2}{z_1(m_2^2 + \p_{23}^2 P_{3g}(z_2))}\cos(2\psi_{12})\right]\,,
\end{align}
This form for the $\cos(2\psi_{12})$ modulation in vacuum for $1\rightarrow 3$ processes with an intermediate gluon is well known in the literature (see e.g. \cite{Shatz:1983hv, Knowles:1987cu,Collins:1987cp, Richardson:2018pvo, Chen:2020adz, Karlberg:2021kwr}). Processes with no intermediate off-shell gluons (e.g. $q\rightarrow qg \rightarrow qgg$) do not exhibit such a dependence.

\subsection{$n$-point medium averages in the large-$N_c$ limit}\label{app:npoint_avs}
The objects $\cK$ and $\cQ$ used in the main text are defined, in the large-$N_c$ limit, as
\begin{subequations}
\begin{align}
	& \cK_{a\rightarrow bg}(\bar t_1, \q_{1}; t_1, \bkappa) = \int_{\u_1,\u_2} e^{i\u_1\cdot\bkappa}e^{-i\u_2\cdot\l}\int_{\u_1}^{\u_2}\cD\u\Exp{i\frac{\omega_a}{2}\int_{t_1}^{\bar t_1}ds\,\dot \u^2}\, \cC^{(3)}_{a\rightarrow bg}(\bar t_1, t_1; \u)\,, \\
	& \cQ_{a\rightarrow bg}(L,\q_1, \q_1; \bar t_1, \bar \bkappa, \l) = \int_{\u_i, \u_f, \bar \u_i, \bar \u_f} e^{i\u_i\cdot\l}e^{-i\bar\u_i\cdot\bar \bkappa}e^{-i(\u_f-\bar\u_f)\cdot\q_1}\nn
	& \times  \int_{\u_i}^{\u_f}\cD\u \int_{\bar\u_i}^{\bar\u_f}\cD\bar\u\Exp{i\frac{\omega_a}{2}\int_{\bar t_1}^Lds\,(\dot \u^2 - \dot{\bar{\u}}^2)}\cC^{(4)}_{a\rightarrow bg}(L,\bar t_1; \u,\bar \u)\,.
\end{align}
\end{subequations}
The $n$-point functions $\cC^{(3)}$ and $\cC^{(4)}$ are given in terms of dipole and quadrupole objects. For the relevant processes, one can write them as
\begin{subequations}
\begin{align}
    & \cC^{(3)}_{q\rightarrow gq}(\bar t, t; \u) = D(\bar t, t; \u)D(\bar t, t; (1-z)\u)\,,\\
    & \cC^{(3)}_{g\rightarrow gg}(\bar t, t; \u) = D(\bar t, t; \u)D(\bar t, t; z\u)D(\bar t, t; (1-z)\u)\,,\\
    & \cC^{(4)}_{q\rightarrow gq}(\bar t, t; \v,\bar \v) = Q(\bar t, t; \v, \bar \v)D(\bar t,t;-(1-z_1)(\v-\bar\v))\,,\\
    & \cC^{(4)}_{g\rightarrow gg}(\bar t, t; \v,\bar \v) = Q(\bar t, t; \v, \bar \v)D(\bar t,t;-(1-z_1)(\v-\bar\v))D(\bar t,t;-z_1(\v-\bar\v))\,,
\end{align}
\end{subequations}
where $U_i^{jk} = U_F^{jk}(\bar t, t; \r_i)$ as defined in Eq.~\eqref{eq:def_wilson_line} and the dipole is defined in Eq.~\eqref{eq:dipole-av}.
The quadrupole is defined as $Q(\bar t, t; \v, \bar \v) = \frac{1}{N_c}\left\langle \Tr\left(U_1^{\dagger}U_2U_3^{\dagger}U_4\right)\right\rangle$, with the relevant differences of coordinates $\r_i$ being defined as 
\begin{subequations}
\begin{align}
	& \r_1-\r_2 = (1-z_1)(\v-\bar\v)\,,\\
	& \r_3-\r_4 = z_1(\v-\bar\v)\,,\\
	& \r_1-\r_4 = \v\, .
\end{align}
\end{subequations}
Keeping this object in its full form results in a quite involved calculation of the $\cC^{(4)}$ function (see e.g.~\cite{Blaizot:2012fh,Apolinario:2014csa, Dominguez:2019ges, Isaksen:2023nlr}). As explained in the main text near Eq.~\eqref{eq:Q_approx}, we simplify it by approximating it as the independent broadening of the two final state partons, which amounts to writing the quadrupole as
\begin{align}\label{eq:indep_broadening}
	Q(\bar t, t; \v,\bar\v) \approx D(\bar t, t; z_1(\v-\bar \v))D(\bar t, t; (1-z_1)(\v-\bar \v)) \equiv \cP(\bar t, t; \v-\bar\v)\,.
\end{align}
With this approximation, everything is given in terms of dipoles, such that $\cK$ and $\cQ$ are simplified to
\begin{subequations}
\begin{align}
	& \cQ_{a\rightarrow bg}(L,\q_1, \q_1; \bar t_1, \bar \bkappa, \l) \equiv \delta^2(\l-\bar\bkappa) \int_\v e^{-i\v\cdot(\q_1-\bar\bkappa)}\cP_{a\rightarrow gb}(L,\bar t_1, \v) \, ,\\
	& \int_{\l}\delta^{2}(\l-\bar\bkappa)\cK_{a\rightarrow bg}(\bar t_1, \l; t_1, \bkappa) \equiv \int_{\u_1,\u_2} e^{i\u_1\cdot\bkappa}e^{-i\u_2\cdot\bar\bkappa}\cK_{a\rightarrow bg}(\bar t_1, \u_2; t_1, \u_1)\,.
\end{align}
\end{subequations}
\bibliographystyle{jhep}
\bibliography{refs.bib}

\end{document}